\documentclass[12pt]{article}
\usepackage[margin=0.95 in]{geometry}
\usepackage{amsmath} 
\usepackage{setspace}
\usepackage{slashed}
\usepackage{comment} 
\usepackage{dsfont}
\usepackage{amssymb,amsfonts}
\usepackage[all]{xy}
\usepackage{graphicx}
\usepackage{relsize}
\usepackage{amsmath}
\usepackage{amssymb}
\usepackage{float}
\usepackage{array}
\usepackage{tikz}
\usepackage{mathtools}
\usepackage{mathrsfs} 
\usepackage{cite}
\usepackage{tikz}
\usepackage{array, multirow}  
\numberwithin{equation}{section}
\newcommand{\cH}{{\mathcal{H}}}

\newcommand{\p}{\partial}

\newcommand{\I}{\mathcal I}
\newcommand{\omegat}{\tilde\omega}
\newcommand{\cO}{{\cal O}}
\newcommand{\be}{\begin{equation}}
\newcommand{\ee}{\end{equation}}

 \newcommand{\Tr}{{\text{Tr}}}
  
\def\bea{\begin{eqnarray}}
\def\eea{\end{eqnarray}}
\allowdisplaybreaks

\numberwithin{equation}{section}
\numberwithin{table}{section}

\usepackage[colorlinks=true, citecolor=black, linkcolor=black, filecolor=black,urlcolor=blue]{hyperref}
\usepackage{amsmath}
\usepackage{amsfonts}
\usepackage{amssymb}
\usepackage{graphicx}
\usepackage{grffile}
\usepackage{subfig}
\usepackage{hyperref}
\usepackage{multirow}
\usepackage{physics}
\usepackage{comment}

\usepackage{braket}
\usepackage{xcolor}
\usepackage{cite}
\usepackage{dsfont}
\usepackage{float}
\usepackage{ulem}
\usepackage{breqn}
\usepackage{tikz}
\def\be{\begin{equation}}
\def\ee{\end{equation}}
\def\bea{\begin{eqnarray}}
\def\eea{\end{eqnarray}}

\newcommand{\Z}{\mathbb{Z}}

\tikzset{
  pics/torus/.style n args={3}{
    code = {
      \providecolor{pgffillcolor}{rgb}{1,1,1}
      \begin{scope}[
          yscale=cos(#3),
          outer torus/.style = {draw,line width/.expanded={\the\dimexpr2\pgflinewidth+#2*2},line join=round},
          inner torus/.style = {draw=pgffillcolor,line width={#2*2}}
        ]
        \draw[outer torus] circle(#1);\draw[inner torus] circle(#1);
        \draw[outer torus] (180:#1) arc (180:360:#1);\draw[inner torus,line cap=round] (180:#1) arc (180:360:#1);
      \end{scope}
    }
  }
}

\begin{document}

\hypersetup{pageanchor=false}
\begin{titlepage}
\vbox{\halign{#\hfil    \cr}}  

\begin{center}

{\Large \bf 
Universal Modular Properties of Generalized \\[2mm] Gibbs Ensembles and
Chiral Deformations
}

\vspace*{5mm}

{\large 
Sujay K. Ashok$^{a,b,c}$, Tanmoy Sengupta$^{d}$,  \\
Adarsh Sudhakar$^{e,b}$,  G\'erard M. T. Watts$^{f}$}

\vspace*{5mm}

$^a$The Institute of Mathematical Sciences, 
		 IV Cross Road,  \\
	C.I.T. Campus, Taramani, Chennai, India 600113

\vspace{.5cm}

$^b$Homi Bhabha National Institute, 
Training School Complex, \\ 
Anushakti Nagar, Mumbai, India 400094

\vspace{.5cm}

$^c$Dipartimento di Fisica, Università di Torino,\\ 
INFN Sezione di Torino,  
Via P. Giuria 1, 10125, Torino, Italy

\vspace{.5cm}

$^d${Chennai Mathematical Institute, H1, SIPCOT IT Park,\\ Siruseri, Kelambakkam 603103, India\\}

\vspace{.5cm}

$^e${Harish-Chandra Research Institute,
Chhatnag Road, \\ Jhunsi,
Allahabad 211019, India}

\vspace{.5cm}

$^f${Department of Mathematics, 
King's College London,\\
Strand, London, WC2R 2LS, United Kingdom}

 \vskip 0.5cm
	{\small
		E-mail:
		\texttt{sashok@imsc.res.in, tanmoys@cmi.ac.in, } \\
  \texttt{\small adarshsudhakar@hri.res.in,  gerard.watts@kcl.ac.uk }
	}
\vspace*{0.5cm}

\end{center}

\begin{abstract} 
 
We study modular properties of conformal field theories perturbed by holomorphic fields. We prove an asymptotic formula for the modular S-transform of a generalized partition function that includes zero modes of higher spin holomorphic currents. The  derivation makes use of  general  properties of torus correlation functions, in particular the Zhu recursion relation. The asymptotic expansion of the modular transformed partition function takes a  universal form that is determined iteratively by the second order pole coefficients in the operator product expansion of the holomorphic currents. We have also found an explicit expression for the multiplicities of terms generated by the iteration.
This proves and generalizes a conjecture regarding the modular transformation properties of generalized Gibbs ensembles.

\end{abstract}

\end{titlepage}

\hypersetup{pageanchor=true}
\setcounter{tocdepth}{2}

{~}\vspace{-1.7cm}
{\begin{spacing}{0.5}
\tableofcontents
\end{spacing}

\section{Introduction}

Modular invariance has played an important role in the study of conformal field theories in two dimensions. The basic idea is very similar to that of the conformal bootstrap \cite{Belavin:1984vu}, but  involves the study of the conformal field theory on the torus. The different channels in the toroidal case depend on the choice of canonical $A$ and $B$ cycles on the torus, and the partition function is computed as a trace of the Hamiltonian evolution operator. The idea of the modular bootstrap \cite{Zamolodchikov:1990ji} is that the partition function itself is independent of the choice of basis of the cycles, which leads to constraints on the spectrum of the evolution operator. For instance, it was shown in \cite{Cardy:1986ie} that modularity   determines the asymptotic behaviour of the partition function, and leads to a universal behaviour for the asymptotic density of states that depends only on the central charge of the conformal field theory. 

In this work we study modular properties of chirally deformed conformal field theories. By this we mean that we deform the conformal field theory by the zero mode of a holomorphic higher spin current. The modular properties of such a deformed theory were first analyzed in \cite{Dijkgraaf:1996iy} for the special case in which the algebra of the deforming operators forms a pre-Lie algebra (this  constrains the first order pole coefficient to be a total derivative). In this work we relax this condition and calculate the modular transformation of a generalized partition function that includes fugacities for the zero modes of arbitrary holomorphic higher spin currents. 

A special case of such chiral deformations is given by the Generalized Gibbs Ensemble.  We recall that a conformal field theory in two dimensions has been shown to possess an integrable structure, with an infinite number of integrals of motion \cite{Sasaki:1987mm, Eguchi:1989hs, Bazhanov:1994ft, Fioravanti:1995cq}. When a theory possesses such higher integrals of motion beyond the Hamiltonian, it is natural to consider generalized partition functions by turning on separate fugacities for each independent conserved charge. In the recent past, the modular properties of such GGEs have been studied in several examples. 

For the Ising model, which is the simplest conformal field theory, the modular properties of the GGE including a finite number of higher spin integrals of motion have been studied in detail in \cite{Downing:2021mfw, Downing:2023lop, Downing:2023lnp}. A similar study was carried out in the Lee-Yang model in \cite{Downing:2024nfb}, in which a single higher spin integral of motion is included in the GGE. The modular properties were studied in an asymptotic expansion in the fugacity for the higher charge. 
It was found that the asymptotic expansion of the modular transform of the GGE is a sum of the zero modes of all the quasi-primary fields of the theory, and cannot be written in terms of the conserved charges of the theory, and actually is not itself a GGE.

A similar analysis has been carried out for a two dimensional conformal field theory with an extended $\mathcal{W}_3$ symmetry algebra \cite{Zamolodchikov:1985wn} \footnote{Conformal field theories with the ${\cal W}_3$ symmetry algebra also have an integrable structure and an infinite number of integrals of motion \cite{Kupershmidt:1989bf, Bazhanov:2001xm,  Ashok:2024zmw, Ashok:2024ygp}.}. The modular properties of the GGE characterized by a single higher spin charge $W_0$, which is the zero mode of a spin three field $W(u)$, was explored in \cite{Downing:2025huv} (see \cite{Gaberdiel:2012yb} for earlier work\footnote{For large central charge, the modular transform of such a GGE was carried out in \cite{Gaberdiel:2012yb} (in an asymptotic expansion in the fugacity) for a conformal field theory with ${\cal W}_{\infty}[\lambda]$ symmetry, and the results matched with the free energy calculation in the gravity dual \cite{Kraus:2011ds}.}). As in the other cases, the modular S-transformation of this ensemble was studied in an asymptotic expansion in the chemical potential for $W_0$. In that work, a very general proposal was made for the S-transformed GGE: the operators whose zero mode appeared in the S-transformation of the GGE were conjectured to satisfy a recursion relation that only depended on the second order pole in the operator product expansion of $W$ with itself. This was checked up to the seventh order in the asymptotic expansion by direct calculation in Verma modules and making use of the properties of the ${\cal W}_3$ algebra \cite{Iles:2013jha, Iles:2014gra}. Recently, for the case of the symplectic free fermion,  the modular properties were derived exactly and found to match with the conjecture \cite{FaisalGerard}.

What is interesting about the proposal in \cite{Downing:2025huv} is its generality; since the result could be stated in terms of the operator product algebra of the higher spin charge, one would expect that such a result could be proven in generality, without having to specialize to any particular algebra. In this work, we show that this is indeed the case. We consider a chiral deformation of a conformal field theory by the zero mode of a spin-$w$ current $W(z)$. 
Without making any restrictions on the operator product algebra, we derive a universal result  for the modular transformation of the deformed theory. We prove that the modular transformation is determined by an iterative relation involving the second order pole in the OPE of $W(z)$ with itself. In the discussion section we show that our result is, in fact, much stronger and that the result holds for a generic holomorphic deformation of the conformal field theory by a local field.  

The organization and layout of the paper will be given at the end of the next section, after introducing some key concepts, presenting a more detailed description of the conjecture, and our approach to its proof. 

\section{Overview of the Conjecture}
\label{prelims}

In this section we shall first of all review some aspects of a two dimensional conformal field theory on a torus. We shall only be concerned with the chiral  sector of the conformal field theory and we review some  properties of torus correlators. We then state and discuss the conjecture that we aim to prove in detail, and we outline a path towards the proof.  

\subsection{Torus Correlators}

We begin with a vertex operator $V(A,z)$  on the plane associated to a state $A$, with conformal dimension $h_A$, and which has the mode expansion
\be 
V(A,z) = \sum_{n\in \mathbb{Z}} A_n\, z^{-n-h_A}~.
\ee 
Here $z$ is a coordinate on the plane. The field on the cylinder is related to the field on the plane by the conformal map  $u\rightarrow z= e^{2\pi i u}$, which is implemented by the unitary map ${\cal U}$:
\be 
 {\cal U} V(A ,u) {\cal U}^{-1} = (2\pi i)^{h_A} V(e^{2\pi i u L_0} \tilde A, e^{2\pi i u}) ~.
\label{operatormap}
\ee 
It is useful to keep in mind that, in the general case, $A$ and $\tilde A$ are distinct. A  constructive way to derive $\tilde A$ can be found in \cite{Gaberdiel:1994fs}.
The torus amplitudes we shall be concerned with have the property that they are single valued under $u\rightarrow u +1$ and $u\rightarrow u+\tau$, where $\tau$ is the complex structure modulus of the torus. Given a representation S of the chiral algebra, one defines the torus amplitudes to be:
\be
\begin{aligned}
\label{toruscorrelator}
   \langle A(z_1) B(z_2)\ldots \rangle_{s, \tau} := \Tr_s\left[V(z_1^{L_0} \tilde A, z_1) V(z_2^{L_0} \tilde B, z_2)\ldots q^{L_0-\frac{c}{24}}\right] ~,
    \end{aligned}
\ee 
where we have defined 
\be 
q=e^{2\pi i \tau}~,\quad \text{and}\quad z_i = e^{2\pi i u_i}~.
\ee 
Here, $L_0-\frac{c}{24}$ is the zero mode of the stress tensor on the cylinder.   
The operators that appear within the trace are rewritten in terms of the vertex operators on the plane. 

Strictly speaking, the operators on the left hand side of equation \eqref{toruscorrelator} should be written as functions of $u_i$. But we will abuse notation and denote the torus amplitudes simply by an angular bracket, with operators denoted as functions of $z_i = e^{2\pi i u_i}$ (we will often also omit the representation label $s$). So whenever operators appear within the angular bracket, it is understood that these are operators on the cylinder. In \cite{zhu:1990, zhu:1996}, it is shown that in a rational conformal field theory, torus correlators involving fields that have definite conformal weight at any  $\tau$ can be expressed in terms of correlators at any other $\tilde \tau$, provided $\tilde \tau$ is related to $\tau$ by a modular transformation. More precisely,
   \begin{multline}  
       (c\tau{+}d)^{-\sum_j h_j} 
       \Tr_r\Big [
       V( e^{\frac{2\pi i u_1}{c\tau+d} L_0}
       \tilde a_1, 
       e^{\frac{2\pi iu_1}{c\tau+d}}) \cdots 
       V( e^{\frac{2\pi i u_n}{c\tau+d} L_0}
       \tilde a_n, e^{\frac{2\pi iu_n}{c\tau+d}}) \, \tilde q^{L_0-\frac{c}{24}} \Big ]\\
       =\sum_s M_r^s \, \Tr_s \Big[ V( e^{2\pi i u_1 L_0 }\tilde a_1, e^{2\pi i u_1}) \cdots V( e^{2\pi i u_n L_0 }\tilde a_n, e^{2\pi i u_n}) \, q^{L_0-\frac{c}{24}}\Big ]~,
\end{multline} 
where the modular transform under consideration is $\tau \to \tilde \tau :=\frac{a\tau+b}{c\tau+d}$ (with $a,b,c,d \in \Z$ and $ad-bc=1$), $\tilde q:=e^{2\pi i\tilde \tau} $ and $M^s_r$ depends only on the representations $s$, $r$ and on the modular transform itself, but not on what the fields $\varphi_1(u_1),\dots \varphi(u_n)$ are. Due to this fact, we can keep the $M$ matrix elements implicit (like we do with the index which denotes the representation we are tracing over), with the understanding that they can be  restored in a straightforward manner.

\subsection{The Conjecture}
\label{conjecture}

We shall consider a holomorphic quasiprimary field $W(z)$ with conformal weight $w$. We do not make any assumption about the chiral algebra of which $W(z)$ is a part. Our goal is to study the modular properties of the generalized partition function: 
\be 
\langle e^{\alpha W_0} \rangle_{\tau} ~.
\label{GGE}
\ee 
Here the zero mode\footnote{Note that this is to be considered as an insertion in a torus correlation function - it is not the zero mode of the field $W$ acting on the Hilbert space in the plane.} $W_0$ is defined as
\be 
\label{zeromodedefn}
W_0 
=
\int_0^1~du~W(u)
= \frac{1}{2\pi i}\oint_A \frac{dz}{z}~W(z)~,
\ee 
where the integral is over the (spatial) $A$-cycle of the torus, which corresponds to the  integral of the periodic $u$-coordinate in the interval $[0,1]$. 
The modular S-transformation is associated to the $\text{SL}(2,\Z)$ element 
\begin{equation}
    S = \begin{pmatrix}
        0 & -1 \\
        1 & 0
    \end{pmatrix}~.
\end{equation}
It acts on the complex structure of the torus as well as the fugacities\footnote{Our conventions are different from those in \cite{Downing:2025huv}, in which $\alpha$ did not transform.}:
\begin{align}
    S:~\tau\rightarrow -\frac{1}{\tau}~, \quad \alpha\rightarrow \frac{\alpha}{\tau^w}~.
    \label{Sonalphatau}
\end{align}
For the specific case when $W_0$ is the zero mode of the spin-3 current in the ${\cal W}_3$ algebra, a proposal was made in \cite{Downing:2025huv} for the modular transform of the asymptotic expansion of the  generating function in \eqref{GGE}. We shall first state a more general version of the  conjecture in which $W_0$ is the zero mode of a  higher spin quasiprimary. 
\begin{align}
    S\left[ \langle e^{\alpha W_0} \rangle_{\tau} \right] :=\langle e^{\frac{\alpha}{\tau^w}\, W_0} \rangle_{-\frac{1}{\tau}}   
    = \langle e^{\alpha  { {\cal W}_0 }} \rangle_{\tau}  ~.
    \label{OGeqn}
\end{align}
Here the second equality is claimed at the level of the asymptotic expansion in $\alpha$, and ${{\cal W}}_0$ corresponds to the zero mode of the field ${\cal W}$, defined by the infinite series:
\be 
\label{calWdefn}
{{\cal W}} = \sum_{n=0}^{\infty}\frac{1}{n!} \left(\frac{\alpha}{4\pi i \tau}\right)^n\, [W^{n+1}]~.
\ee 
Each term $[W^{n}]$ appearing on the RHS is a composite operator of conformal weight $n w -2 (n-1)$ and the conjecture is that it is obtained recursively: 
\be 
\label{Wnrecursionv2}
[W^{n+1}] = (WW)_2\frac{\partial}{\partial W}[W^n]~.
\ee 
where the product $(AB)_2$ is the second order pole in the OPE of $A$ and $B$, with $[W] = W$. We shall formally define the variational derivative  in Section \ref{provingconjecture}, but it should be understood as a shorthand for the replacement operation, in which each $W$-operator is replaced by the operator $(WW)_2$. For instance
\begin{align}
[W^2] = (WW)_2~,\quad [W^3] = ((WW)_2W)_2+ (W(WW)_2)_2~, 
\end{align} 
and so on. 

Now that we have stated the conjecture, let us analyze the two sides of the equation in greater detail. It is important to stress that in the equality  between the two sides in \eqref{OGeqn}, both sides are expanded as a power series expansion in $\alpha$. The left hand side of that equation is simply given by\footnote{From here onwards, we omit the $\tau$ subscript for the angular brackets.}
\be 
 S\left[ \langle e^{\alpha W_0} \rangle_{\tau} \right] = \sum_{n=0}^{\infty} \frac{\alpha^n}{n!} \frac{1}{\tau^{nw}} S\Big[\langle W_0^n \rangle_{\tau}\Big]~.
\label{LHSexpansion}
\ee 
Note that the S-transform acts on the $\alpha$-parameter as in \eqref{Sonalphatau}.
From general considerations, it is known that $\langle W_0^n\rangle$ is a quasi-modular form of weight $wn$ and depth $(n-1)$ \cite{Dijkgraaf:1996iy}. This means that it can be expanded as a polynomial in the Eisenstein series $E_2(\tau)$, with modular forms as coefficients: 
\begin{equation}
    \langle W_0^n\rangle_{\tau} = \mathlarger{\mathlarger{\sum}}_{i=0}^{n-1} \alpha_i f_{wn-2i}(\tau)E_{2}^{i}(\tau),
\end{equation}
Here $f_{wn-2i}(\tau)$ is modular form of weight $wn-2i$. The modular transformation of such an expression is given by
\begin{equation}
    \begin{aligned}
       S\Big[ \langle W_0^n\rangle_{\tau}\Big]=&\mathlarger{\mathlarger{\sum}}_{i=0}^{n-1}\alpha_i \tau^
       {wn-2i}f_{wn-2i}(\tau)(\tau^2E_{2}(\tau)  -\frac{6i}{\pi}\tau)^{i} \\
=&\mathlarger{\mathlarger{\sum}}_{i=0}^{n-1}\sum_{k=0}^{i}\alpha^{\prime}_{i,k} \tau^
       {wn-2i}f_{wn-2i}(\tau)(\tau^2E_{2}(\tau))^k  (\tau)^{i-k} \\
=&\mathlarger{\mathlarger{\sum}}_{i=0}^{n-1}\sum_{k=0}^{i}\alpha^{\prime}_{i,k} \tau^
       {wn-i+k}f_{wn-2i}(\tau)(E_{2}(\tau))^k~.  \\
    \end{aligned}
    \label{Stransformgeneral}
\end{equation} 
At the moment, our main takeaway is the form of the S-transform on the right hand side, which follows from general considerations. We have a polynomial in $\tau$, with coefficients given by linear combinations of quasi-modular forms.  

Let us see how this form of the S-transform can arise from the point of view of the torus correlators of the zero modes. Given the definition of the zero mode in \eqref{zeromodedefn}, 
the correlator of $n$ zero modes can be written as 
\be 
\langle W_0^n \rangle_{\tau} =\frac{1}{(2\pi i)^n} \oint_A \prod_{i=1}^n \frac{dz_i}{z_i} \langle W(z_1)\ldots W(z_n) \rangle_{\tau}~. 
\label{Aperiodintegrals}
\ee 
where each integral is over the $A$-cycle of the torus.  
Let us consider  the $n$-point correlator  and recall from equation \eqref{toruscorrelator} its expression as the trace:
\be 
\langle W(z_1) \ldots W(z_n) \rangle_{\tau} =  \Tr \left[V(z_1^{L_0}\tilde W, z_1)\ldots V(z_n^{L_0}\tilde W, z_n) \, q^{L_0-\frac{c}{24}} \right]~.
\label{correlatortransform}
\ee 
Under modular transformation, it has been proven by Zhu in \cite{zhu:1996} (see Theorem 5.3.2) that such a torus correlator transforms homogeneously as long as the state corresponding to $W$ transform homogeneously under the scaling transformations on the plane.  
So, for the  local field
\footnote{Although Theorem 5.3.2 in \cite{zhu:1996} is proven for primaries, the postscript of the theorem clearly states the stronger result quoted here.} $W(z)$ of definite weight $w$, under the S-transformation, the correlator picks up a factor $\tau^{nw}$. So, for the integrated correlator, we obtain 
\be
\begin{aligned}
\frac{1}{(2\pi i)^n}\oint_A \prod_{i=1}^n \frac{dz_i}{z_i} \langle W(z_1) \ldots W(z_n) \rangle_{-\frac{1}{\tau}} &=\frac{1}{(2\pi i)^n}\oint_A \prod_{i=1}^n \frac{dz_i}{z_i}\tau^{nw}\, \langle W(z_1^{\tau}) \ldots W(z_n^{\tau}) \rangle_{\tau} \\ 
\end{aligned}
\ee 
Changing variables within the integrated correlator from $z$ to $z'=z^{\tau}$, as explained in \cite{Gaberdiel:2012yb}, we  pick up a factor of $\tau^{-n}$ from the transformation of the measure, and we effectively obtain an integral over the $B$-period of the torus with modular parameter $\tau$, leading to the modular transformation: 
\begin{equation}
    \begin{aligned}
        S\Big[\langle W_0^n \rangle_{\tau}\Big] &= \frac{\tau^{n(w-1)}}{(2\pi i)^n} \oint_B \prod_{i=1}^n \frac{dz'_i}{z'_i} \langle W(z'_1) \ldots W(z'_n) \rangle_{\tau} ~,
    \end{aligned}
    \label{Bperiodintergrals}
\end{equation}
where the $B$-period integral corresponds to the integral of the periodic $u'$-coordinate in the interval $[0,\tau]$, or in the $z'$-coordinate, from $1$ to $q$. This is illustrated in figure \ref{fig: Figure from 2021}. For more details, and in particular how the different chiral sectors contribute to a fully modular invariant expression, we refer the reader to \cite{Downing:2021mfw}.\\

\begin{figure}[htb]
\[
\begin{array}{cccccccccc}
&&
\hfill\mbox{\Large$\llcorner$}\raisebox{1mm}{\kern -1.4mm{$u$}}
&&
\hfill\mbox{\Large$\llcorner$}\raisebox{1mm}{\kern -1.4mm{$z$}}
&&
\hfill\mbox{\Large$\llcorner$}\raisebox{1mm}{\kern -1.4mm{$u'$}}
&&
\hfill\mbox{\Large$\llcorner$}\raisebox{1mm}{\kern -1.4mm{$z'$}}
\\[-3mm]
\begin{tikzpicture}[baseline=-1.1em]
 \pic{torus={1cm}{2.8mm}{70}}
      ;
\begin{scope}[yshift=0em,yscale = 0.35]
    \draw[red] (1.28,0) arc (0:-180:1.28);
    \draw[teal] (-0.2,-.16) arc (165:195:3.1);
    \draw[red] (0.2,-1.03)--(0.3,-1.23)--(0.2,-1.43);
    \draw[teal] (-0.38,-.9)--(-0.29,-.65)--(-0.22,-.95);
\end{scope}
\end{tikzpicture}
&
\raisebox{4mm}{\hbox{$\equiv$}}
&
\begin{tikzpicture}[baseline=1.3em]
\draw (0,0.2) -- (1.5,0.2)  node [right] {$\scriptstyle 1$};
\fill[black] (0,0.2) circle (0.05);
\fill[black] (-0.4,1.7) circle (0.05) node [above] {$\scriptstyle-1/\tau$};
\draw (0,0.2) -- (-0.4,1.7);
\draw[teal,->] (0.15,.2) -- (0.017,.7);
\draw[teal] (0.017,0.7) -- (-0.25,1.7);
\draw[red] (1.45,.35) -- (.75,.35);
\draw[red,->] (-0.04,.35) -- (.75,.35);
\draw (1.5,0.2) -- (1.1,1.7) -- (-0.4,1.7);
\draw (0.4,0.1)--(0.6,0.3);
\draw (0.06,1.6)--(0.26,1.8);
\draw (-.3,0.85)--(-.1,1.05);
\draw (-.36,0.85)--(-.16,1.05);
\draw (1.5-.3,0.85)--(1.5-.1,1.05);
\draw (1.5-.36,0.85)--(1.5-.16,1.05);
\end{tikzpicture}
&
\raisebox{4mm}{\hbox{$\equiv$}}
&
\begin{tikzpicture}[baseline=2em]
\begin{scope}[yshift=3em]
\draw[red,->] (0.7,0) arc (0:180:0.7);
\draw[red] (-0.7,0) arc (180:360:0.7);
\draw[teal] (0.8,0) arc (-30:-110:0.6);
\draw[teal] (0.55,-0.33) -- (0.45,-0.3) -- (0.48,-0.18);
\draw (0,0) circle (0.8);
\draw (0,0) circle (0.3);
\fill[black] (0.8,0) circle (0.05) node [right] {$\scriptstyle 1$};;
\fill[black] (0.1,-0.28) circle (0.05) node [above] {$\scriptstyle \hat q$};;
\draw (-0.38,0.85)--(-0.26,0.65);
\draw (-0.27,0.3)--(-0.15,0.1);
\end{scope}
\end{tikzpicture}
&
\raisebox{4mm}{\hbox{$\equiv$}}
&
\begin{tikzpicture}[baseline=2em]
\draw (0,0.2) -- (1.7,0.2)  node [right] {$\scriptstyle 1$};
\fill[black] (0,0.2) circle (0.05);
\fill[black] (0.8,2.2) circle (0.05) node [above] {$\scriptstyle\tau$};
\draw (0,0.2) -- (0.8,2.2);
\draw[red,->] (0.15,0.2) -- (0.35,0.7);
\draw[red] (0.35,0.7) -- (0.95,2.2);
\draw[teal] (.05,0.35) -- (1.1,0.35);
\draw[teal,->] (1.75,0.35) -- (1.1,0.35);
\draw (1.7,0.2) -- (2.5,2.2) -- (0.8,2.2);
\draw (0.79,0.3)--(0.91,0.1);
\draw (0.74,0.3)--(0.86,0.1);
\draw (1.59,2.3)--(1.76,2.1);
\draw (1.54,2.3)--(1.71,2.1);
\draw (0.3,1.3)--(0.5,1.1);
\draw (2.02,1.3)--(2.22,1.1);
\end{tikzpicture}
&
\raisebox{4mm}{\hbox{$\equiv$}}
&
\begin{tikzpicture}[baseline=2em]
\begin{scope}[yshift=3em]
\draw (0,0) circle (0.8);
\draw (0,0) circle (0.3);
\fill[black] (0.8,0) circle (0.05) node [right] {$\scriptstyle 1$};;
\fill[black] (0.1,0.28) circle (0.05) node [above] {$\scriptstyle q$};;
\draw (-0.38,0.85)--(-0.26,0.65);
\draw (-0.27,0.3)--(-0.15,0.1);
\draw (-0.43,0.85)--(-0.31,0.65);
\draw (-0.22,0.3)--(-0.1,0.1);
\draw[teal] (0.4,0) arc (0:180:0.4);
\draw[teal,<-] (-0.4,0) arc (180:360:0.4);
\draw[red] (0.8,0) arc (30:110:0.6);
\draw[red
] (0.55,0.33) -- (0.45,0.3) -- (0.48,0.18);
\end{scope}
\end{tikzpicture}
\\
(a) && (b) && (c) && (d) && (e) \end{array}
\]
\caption{(a) A torus with $A$ and $B$ cycles realized as a quotient of the plane in two ways (b) and
 (d), with $u' = u\tau + 1 \equiv u\tau$, and correspondingly as an annulus in two ways with (c)  $z=\exp(2\pi i u)$, 
 and (e) $z' = \exp(2\pi i u')$.
 The initial integration over $u$ along the (red) spatial A cycle from 0 to 1 in (b) as in equation \eqref{Aperiodintegrals} becomes an integration from 0 to $\tau$ in (d) and over $z'$ from 1 to $q=\exp(2\pi i \tau)$ in (e) as in equation \eqref{Bperiodintergrals}.}
\label{fig: Figure from 2021}
\end{figure}
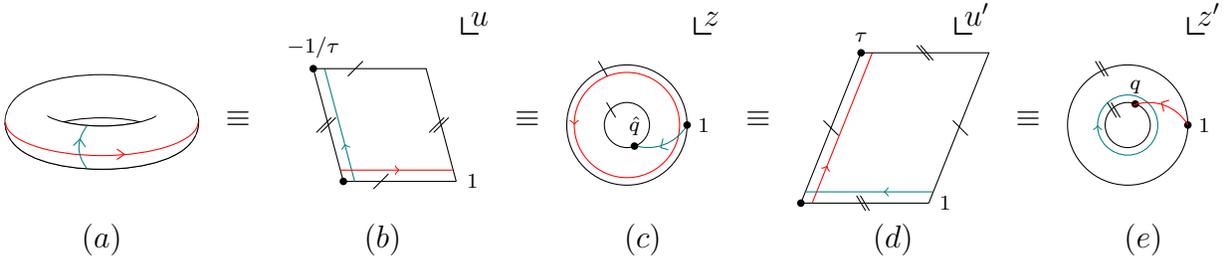
The exact nature of the multidimensional contour will be discussed in detail in Section \ref{Zhutostart}, and we shall relabel $z'$ by $z$ to avoid clutter. Thus the asymptotic expansion of the left hand side of the conjecture in \eqref{OGeqn} is given by (see equation \eqref{LHSexpansion})
\begin{align} 
 S\left[\langle e^{\alpha W_0} \rangle_{\tau}\right]  &= \sum_{n=0}^{\infty} \frac{\alpha^n}{n!} \frac{1}{\tau^{nw}} S\Big[\langle W_0^n \rangle_{\tau}\Big] \cr
 &= \sum_{n=0}^{\infty}\frac{\alpha^n}{n!} \frac{1}{(2\pi i \tau)^n} \oint_B \prod_{i=1}^n \frac{dz_i}{z_i} \langle W(z_1) \ldots W(z_n) \rangle_{\tau} ~.
\end{align} 

The right hand side of the conjecture is more complicated, as the field ${\cal W}$ itself depends on $\alpha$; the exponent in the RHS in \eqref{OGeqn} is given by
\be 
\alpha  {{\cal W}} = \alpha  W + \frac{\alpha^2}{(4\pi i \tau)}  [W^2] + \frac{1}{2!}\frac{\alpha^3}{(4\pi i \tau)^2}  [W^3]+ \frac{1}{3!}\frac{\alpha^4}{(4\pi i\tau)^3}  [W^4]+\ldots  \ldots
\ee 
Taking the exponent of the zero mode of this operator, and doing an (asymptotic) expansion for small $\alpha$, we obtain:
\begin{equation}
\begin{aligned}
    \langle e^{\alpha {{\cal W}}_0} &\rangle_{\tau} = \langle 1\rangle + \alpha\langle W_0\rangle_{\tau} + \frac{\alpha^2}{2!(2\pi i\tau)^2}\Big[(2\pi i\tau)^2\langle W_0^2\rangle_{\tau} +(2\pi i\tau)\langle [{ W}^2]_0\rangle_{\tau} \Big]\\
    &+ \frac{\alpha^3}{3!(2\pi i\tau)^3} \Big[(2\pi i\tau)^3\langle W_0^3\rangle_{\tau} +   3(2\pi i\tau)^2\langle W_0\, [{ W}^2]_0 \rangle_{\tau}+ \frac{3}{4}(2\pi i\tau)\langle[{ W}^3]_0\rangle_{\tau} \Big]\\
    &+  \frac{\alpha^4}{4!(2\pi i\tau)^4} \Big[(2\pi i\tau)^4\langle W_0^4\rangle_{\tau}+ 6(2\pi i\tau)^3  \langle W_0^2 [W^2]_0\rangle_{\tau} +\ldots 
    + \frac{1}{2}(2\pi i\tau)\langle[{ W}^4]_0\rangle_{\tau} \Big]+ \ldots 
\end{aligned}
\label{RHSexpansion}
\end{equation}
By comparing the coefficients in the asymptotic expansion in $\alpha$ (equations \eqref{LHSexpansion} and \eqref{RHSexpansion}) we see that the integrated $n$-point correlator of the $W$s is equal to the terms within the brackets. The zero modes $[W^n]_0$ typically do not commute with one another, and due to the exponentiation (which involves only the traces $\langle \mathcal W_0^n\rangle$) one finds that all traces are symmetrized with respect to the zero modes (see equation (3.4) of \cite{Downing:2025huv} for the first instance where the symmetrization occurs.). 
This symmetrization ensures that the correlators turn out to be quasi-modular, as would be necessitated by the general expression in \eqref{Stransformgeneral}. 

At this point, we make an observation: up to an overall constant, the composite operators $[W^n]$ appearing in the field ${\cal W}$ can be identified from the $\cO(\tau)$ term appearing in the integrated correlator\footnote{The integration limits indicate that we are integrating over the $B$-cycle of the torus.}:
\be 
\label{ordertauintegcorr}
\int_1^q \prod_{i=1}^n \frac{dz_i}{z_i} \langle W(z_1) \ldots W(z_n) \rangle_{\tau} = \frac{n}{2^{n-1}}\langle [{ {W}}^n] \rangle_{\tau}\, (2\pi i\tau) + \cO(\tau^2) 
\ee 

Thus, given the form of the conjecture about the composite operators $[W^n]$ in \eqref{Wnrecursionv2}, our goal will be to isolate the $\cO(\tau)$ coefficient of the integrated $n$-point correlator and to derive its recursive properties. The main tool that allows us to lay bare the recursive nature of the integrated correlator is a recursion relation  satisfied by torus correlators, first derived by Zhu \cite{zhu:1990, zhu:1996}. 

This paper is organized as follows. In Section \ref{squaremodes} we show the relationship between operator product coefficients on the cylinder and the action of square mode operators, that naturally make their appearance in the Zhu recursion relation. This allows us to restate the conjecture to be proven in terms of square modes. In Section \ref{Zhutostart} we review the Zhu recursion relation and go on to derive some general results regarding integrated correlators. We then describe in detail the application of the Zhu recursion to the particular integrated $n$-point correlator of interest in Section \ref{intcorr}. The main result of this section is a recursive relation, in which we write the integrated $n$-point correlator in terms of lower point ones. Finally, we prove in Section \ref{provingconjecture} that, as a consequence of this recursion, the  coefficient of the linear term in $\tau$ of the integrated correlator satisfies precisely the recursion relation expected from the conjecture. We end with a detailed discussion in  Section \ref{conclusions}. First of all we summarize and extend our results:  we derive the modular transformation of partition functions associated to generic chiral deformations of conformal field theories. We also work out simple applications of our general result and make contact with the earlier work  of Dijkgraaf \cite{Dijkgraaf:1996iy}. Technical details are collected in the Appendices.

\section{Square Modes and OPE Coefficients}
\label{squaremodes}

In this section, we shall 
first review the square modes associated to vertex operators \cite{zhu:1996} and then show how they are related to operator product expansions of fields on the cylinder. 
As discussed previously, the vertex operator $V(A,z)$  on the plane associated to a state $A$ and conformal dimension $h_A$, with the mode expansion
\be 
V(A,z) = \sum_{n\in \mathbb{Z}} A_n\, z^{-n-h_A}~.
\ee 
The map between the cylinder and the plane is accomplished  by the unitary map ${\cal U}$:
\be 
 {\cal U} V(A ,u) {\cal U}^{-1} = (2\pi i)^{h_A} V(e^{2\pi i u L_0} \tilde A, e^{2\pi i u}) ~.
\label{operatormap1}
\ee 
This equation defines the operator $\tilde A$. Following \cite{zhu:1996} we define the vertex operator:
\be
V[A,u]:=V(e^{2\pi i u L_0} \tilde A, e^{2\pi i u} -1)~.
\ee 
The formal expansion of $V[A,u]$ for small $u$ defines the square modes:
\be 
V[A,u] = \sum_{n\in \mathbb{Z}}\frac{\tilde A[n]}{u^{n+1}}~.
\ee

\subsection{The  Operator Product Expansion}
\label{opeandsquaremodes}

The action of square modes on operators is closely related to OPE coefficients. Consider the OPE of two local operators on the plane:
\be
V(A,z)V(B,w) = \sum_{n}\frac{(AB)_n(w)}{(z-w)^n}~.
\label{OPEAandB}
\ee 
Using the plane to cylinder map, we obtain:
\begin{align}
V(e^{2\pi i u L_0} \tilde A, e^{2\pi i u}) V(    e^{2\pi i v L_0} \tilde B, e^{2\pi i v}) &= \frac{1}{(2\pi i)^{h_A+h_B}}{\cal U} V(A, u){\cal U}^{-1}{\cal U}V(B,v){\cal U}^{-1} \cr
&=  
 \sum_{n}\frac{1}{(2\pi i)^{h_A+h_B }}\frac{{\cal U}(AB)_n(v){\cal U}^{-1}}{(u-v)^n} \cr
 &=  \sum_{n}\frac{1}{(2\pi i)^{h_A+h_B-h_{(AB)_n} }}\frac{V( e^{2\pi i vL_0}\widetilde{(AB)}_n , e^{2\pi i v})}{(u-v)^n}
  ~.   
  \label{OPEcylinderfromplane}
\end{align}
 Consider the product of two vertex operators on the cylinder; we use the duality property in a meromorphic conformal field theory (see \cite{Goddard:1989dp, Gaberdiel:1999mc}) for the same product of vertex operators: 
\begin{align}
V(e^{2\pi i u L_0} \tilde A, e^{2\pi i u}) &V(    e^{2\pi i v L_0} \tilde B, e^{2\pi i v}) =V(V(e^{2\pi i u L_0} \tilde A, e^{2\pi iu} - e^{2\pi iv})e^{2\pi i v L_0}\tilde B,e^{2\pi iv}  ) \cr
&= V(e^{2\pi i v L_0} e^{-2\pi i v L_0}V(e^{2\pi i u L_0} \tilde A, e^{2\pi iu} - e^{2\pi iv})e^{2\pi i v L_0}\tilde B,e^{2\pi iv}  ) \cr
&=V(e^{2\pi i v L_0} V(e^{2\pi i (u-v) L_0} \tilde A, e^{2\pi i(u-v)} - 1) \tilde B,e^{2\pi iv}  ) \cr
&= \sum_{n\in \mathbb{Z}} \frac{ V(e^{2\pi i v L_0} \tilde A[n]\tilde B, e^{2\pi i v})}{(u-v)^{n+1}}~. 
\label{VVduality}
\end{align} 
In the third equality we have used
\be 
e^{-2\pi i v L_0}V(C, e^{2\pi i w}) e^{2\pi i vL_0} = V(e^{-2\pi i v L_0}C, e^{2\pi i (w-v)})~.
\end{equation}
By comparing \eqref{VVduality}  with \eqref{OPEcylinderfromplane}, we see that the operator that appears as the coefficient of the $n$-th order pole is related to the square mode action by the relation
\be 
\widetilde{(AB)}_n = (2\pi i)^{n}\, (\tilde A[n-1]\tilde{B})~.
\label{nthorderpole}
\ee 
Given the definition of the correlators on the torus in \eqref{toruscorrelator}, what this means is that one should think of square mode action on operators as the OPE coefficient\footnote{We choose to write our expressions in terms of square mode actions on operators as this is what naturally arises in the Zhu recursion relation, as we shall see in the subsequent analysis.}:
\be 
(2\pi i)^{n}\langle \ldots (A[n-1]B)(z) \ldots \rangle_{\tau} = \langle \ldots (AB)_n(z) \ldots \rangle_{\tau}
\ee 

\subsubsection{Restating the Conjecture} 

Given the relation between the OPE coefficients and the square modes, the conjecture for the composite operator appearing in the modular S-transform of the zero mode expectation value can be rewritten in the form:
\be 
\langle [W^{n+1}]\rangle_{\tau} = (2\pi i)^2\, \langle (W[1]W)\frac{\partial}{\partial W}[W^n] \rangle_{\tau} ~.
\label{Wnrecursionv3}
\ee 
This is the relation that we shall aim to prove.

\section{The Integrated Correlator}
\label{Zhutostart}

In this section we begin with the integral over the $B$-periods that occur in the modular transform of the zero mode expectation values, and apply the Zhu recursion to the $n$-point correlator. We are interested in extracting the coefficient of linear-in-$\tau$ term of the integrated correlator. Since we shall make extensive use of the recursive procedure, we shall briefly review the generalized recursion relations satisfied by torus correlators. 

\subsection{The Zhu Recursion}

An $n$-point torus correlator satisfies the recursion relation \cite{zhu:1990, zhu:1996}
\be 
\begin{aligned}
\langle X_1(z_1)\ldots X_n(z_n) \rangle = & \langle (X_{1})_0\,   X_2(z_2)\ldots X_n(z_n) \rangle \\ 
&+ \sum_{j=2}^n \sum_{m\ge 0}{\cal P}_{m+1}\Big(\frac{z_j}{z_1} \Big) \langle X_2(z_2) \ldots (X_1[m]X_j)(z_j)\ldots X_n(z_n) \rangle ~.
\end{aligned}
\label{basicZhu}
\ee 
From this section onwards, we shall omit the $\tau$-subscript in the angular brackets to avoid cluttering the formulae.
Here $(X_{1})_0$ denotes the zero mode of the field $X_1(z)$, and ${\cal P}_{m+1}(z)$ are the generalized Weierstrass functions, which we define by the power series \cite{zhu:1996}:
\be 
{\cal P}_{m+1}(z) = \frac{(2\pi i)^{m+1}}{m!} \sum_{n\in\mathbb{Z}} \frac{n^m z^n}{1-q^n} ~. 
\ee
As one can see, the right hand side of  equation \eqref{basicZhu} is a lower point correlator and, in principle, it is possible to continue this until one obtains a linear combination of correlators involving zero modes of composite operators. Of course, one does find a new kind of correlator on the right hand side, in which there is a zero-mode insertion. So, in order to proceed, one needs a generalized recursion -- derived in \cite{Gaberdiel:2012yb}-- and which is applicable to such correlators. For our purposes, we shall only need the generalized recursion involving a single zero mode insertion; since we are only interested in the coefficient of the $\cO(\tau)$ term of integrated correlators, we will not need the recursion for multiple zero mode insertions, since the integrated correlator would be at least $\cO(\tau^2)$. The generalized Zhu recursion that we shall use in this work is given by
\cite{Gaberdiel:2008pr, Gaberdiel:2012yb}:
\be 
\begin{aligned}
    \langle (X_1)_0\, X_2(z_2)\ldots &X_n(z_n)\rangle =  \langle (X_1)_0\,  (X_{2})_0\, X_3(z_3)\ldots X_n(z_n)\rangle\\
    &+\sum_{j=3}^n \sum_{m_j\ge 0}{\cal P}_{m_j+1}\Big(\frac{z_{j}}{z_2}\Big)\langle (X_1)_0\,   X_3(z_3)\ldots (X_2[m_j]X_j)(z_j) \ldots X_n(z_n)\rangle\\
    &-\sum_{j=3}^n \sum_{m_j\ge 0}g^1_{m_j+1}\Big(\frac{z_{j}}{z_2}\Big)\langle  X_3(z_3)\ldots ((X_1[0]X_2)[m_j]X_j)(z_j) \ldots X_n(z_n)\rangle 
    \label{zeromodeZhu}
\end{aligned}
\ee 
We shall omit the first term on the right hand side in this work since, upon integration over $z_1$ and $z_2$, it would contribute at $\cO(\tau^2)$. 
In the equation, $g^1_{m+1}(z)$ is the derivative of the Weierstrass function:
\be 
g^1_{m+1}(z) =\frac{ (2\pi i) }{m} \partial_{\tau} {\cal P}_m(z)~.
\ee 

\subsubsection{The Integrated Correlator}

In the integrated correlator of interest, all operators are identical and we denote this by $f_n$:
\be 
f_n =  \int d\mu_{1n} \langle  W(z_1)\ldots W(z_n)\rangle ~.
\ee 
For the first step of the Zhu recursion, we use the Zhu recursion in \eqref{basicZhu}:
\begin{equation}
    \begin{aligned}
    f_n =& \int d\mu_{1n} \langle W_0\, W(z_2)\ldots W(z_n)\rangle \\
     &+\sum_{j=2}^{n}\int d\mu_{1n} \sum_{m\ge0}{\cal P}_{m+1}\Big(\frac{z_j}{z_1}\Big) \langle W(z_2)\ldots (W[m]W)(z_j)\ldots W(z_n)\rangle~.
    \end{aligned}
    \label{ZhuWnfirststep}
\end{equation}
We have defined the measure \cite{Gaberdiel:2012yb}:
\be 
d\mu_{kn} =  \int_1^q \frac{dz_n}{z_n}  \ldots\int_{\epsilon_2}^{\epsilon_2 q} \frac{dz_{k+1}}{z_{k+1}}\int_{\epsilon_1}^{\epsilon_1 q}\frac{dz_k}{z_k}\,~.
\ee 
The integration starts with $z_k$ and goes up to $z_n$. The $\epsilon_i$ are phases chosen such that the angles arg$(\epsilon_i) < \text{arg}(\epsilon_{i+1})$. For more details about the contours, we refer to \cite{Gaberdiel:2012yb}.

\subsubsection{Order of Integration}
\label{orderofintegrals}

A very useful result that we shall prove below is to compute the result of exchanging the order of integration. Consider a term of the form :
\begin{equation}
    \int_1^q  \frac{dz_2}{z_2} \int_\epsilon^{\epsilon q} \frac{dz_{1}}{z_{1}}   \langle X_1(z_1) X_2(z_2) \dots  X_n(z_n)\rangle~, 
\end{equation}
where $\epsilon$ is some phase, such that $\text{arg}\, \epsilon < 0$. We assume that the arguments of $z_i$ for $i>2$ do not lie in $(\text{arg}\, \epsilon,0)$. As with usual contour deformation arguments, we note that the integrand is singular when $z_1=z_2$, and moving the $z_1$ integration contour across this singularity means that we pick up terms coming from the residue at $z_1=z_2$ : 
\begin{equation}
\begin{aligned}
   \int_1^q  \frac{dz_2}{z_2} \int_\epsilon^{\epsilon q} \frac{dz_1}{z_1}   \ \langle X_1(z_1) X_2(z_2) &\dots X_n(z_n)\rangle 
    -  \int_\epsilon^{\epsilon q} \frac{dz_1}{z_1}  \int_1^q  \frac{dz_2}{z_2} \ \langle X_1(z_1) X_2(z_2) \dots X_n(z_n)\rangle \\ 
    &= \int_1^q  \frac{dz_2}{z_2} \oint_{z_2} \frac{dz_1}{z_1}   \ \langle  X_1(z_1) X_2(z_2)\dots X_n(z_n) \rangle \\ 
    &= 2\pi i \int_1^q \frac{dz_2}{z_2} \oint_{u_2} du_1 \  \langle  X_1(z_1) X_2(z_2)\dots X_n(z_n) \rangle~, 
\end{aligned}   
\end{equation}
where we used the fact that $z_2$ lies on the line from $1$  to $q$. Within the correlator, now the relevant bit is the OPE expansion of the field at $z_1$ with the field at $z_2$:
\begin{equation}
\begin{gathered}
  \sum_{k=0}^\infty (2\pi i)^{k+2} \int_1^q \frac{dz_2}{z_2} \frac{1}{2\pi i}\oint_{u_2} \frac{du_1}{(u_1-u_2)^{k+1}} \  \langle (X_1[k]X_2)(z_2) \dots X_n(z_n) \rangle~.
\end{gathered}   
\end{equation}
Evaluating the $u_1$ integral, we find :
\begin{equation}
\begin{gathered}
  (2\pi i)^2 \int_1^q \frac{dz_2}{z_2} \  \langle (X_1[0]X_2)(z_2) \dots X_n(z_n) \rangle~.
\end{gathered}   
\end{equation}
This is the additional term one gets on commuting $X_2(z_2)$ to the left of $X_1(z_1)$. 

\subsection{A Simplified Recursion}

Let us apply this result to the first level Zhu recursion done on the integrated $n$-point correlator (see equation \eqref{ZhuWnfirststep}) in order to simplify our analysis: 
\begin{equation}
    \begin{aligned}
    f_n = & \int d\mu_{1n} \langle W_0\, W(z_2)\ldots W(z_n)\rangle \\
     &+\sum_{j=2}^{n}\int d\mu_{1n} \sum_{m\ge0}{\cal P}_{m+1}\Big(\frac{z_j}{z_1}\Big) \langle W(z_2)\ldots (W[m]W)(z_j)\ldots W(z_n)\rangle~. 
    \end{aligned}
\end{equation}
Consider the $j$th term in the sum, and move the composite operator $W[m]W(z_{j})$ all the way to the left. 
By using the result we obtained for changing the order of integration, we see that, when one commutes the  $W[m]W$ operator onto the $(j-1)$th position, one generates an extra term, given by the integrated $(n-2)$-point correlator of the form (in the measure, we skip $z_j$ and shifted the labels down by one):
\be 
\int d\mu_{1,\slashed{j},n-1} \sum_{m\ge 0} {\cal P}_{m+1}\Big(\frac{z_{j-1}}{z_1}\Big) \langle W(z_2)\ldots W(z_{j-2}) (W[0](W[m]W))(z_{j-1})\ldots W(z_{n-1})\rangle 
\ee 
One can now use the result that the integrated correlator with all $W(z_i)$ and one operator of the form $(W[0]X)(z)$ vanishes:
\be 
\int d\mu_{2,\slashed{j},n-1} ~h_m(z_{j-1}) ~\langle W(z_2)\ldots W(z_{j-2})(W[0]X)(z_{j-1}) \ldots W(z_{n-1})\rangle =0 ~.
\label{WOXlemmaintext}
\ee 
In this expression, the integral over $z_1$ has been done and independently of what function $h_m$ we get, the integrated correlator vanishes.  This is proved as Lemma 2 (see equation \eqref{lemma2}) in Appendix \ref{realLemma}. 
Thus, all the composite operators can be moved to the left without generating any addition terms.
Thus, using the commutativity of the operators, and after performing the first step of the Zhu recursion, the integrated $n$-point correlator can be written in the simplified form\footnote{A similar form of the recursion is also used in \cite{Gaberdiel:2012yb}. In that case, the composite operator is written at $z_n$.}: 
\begin{equation}
    \begin{aligned}
    f_n=& \int d\mu_{1n} \langle W_0\, W(z_2)\ldots W(z_n)\rangle \\
     &+(n-1)\int d\mu_{1n} \sum_{m\ge0}{\cal P}_{m+1}\Big(\frac{z_2}{z_1}\Big) \langle (W[m]W)(z_2)W(z_3)\ldots W(z_n)\rangle~.
    \end{aligned}
    \label{originalZhu}
\end{equation}
This will be the starting point for recursing the integrated correlator. 

\section{Recursion of the Integrated Correlator} 
\label{intcorr}

In this section we shall write the integrated $n$-point correlator as a sum of lower point integrated correlators by repeated use of the Zhu recursion relation. The final result is written in a compact form in \eqref{npointaslowerpoint}. It will be a rather long and technical analysis to get to that point, but there are some recurring ideas that help in our analysis. We list a few of these below to help navigate the section. 

The first simplification arises from the constraints of vanishing integrals; as a consequence, the summation that appears in the Zhu recursion is often truncated  to just one or two terms. Sometimes it is also useful to rewrite a complicated integral in terms of a sum of simpler ones. Such useful integrals have been listed in Appendix \ref{listofintegrals}. Secondly, we shall often use operator identities that follow from the Jacobi identity, or from flipping the order of operators within composite operators (for instance, $A[m]B$ being rewritten in terms of $B[m]A$ etc.). Lastly, we shall repeatedly use three ``vanishing lemmas" for integrated correlators; one of them we have already encountered in \eqref{WOXlemmaintext}, and there are two others that have been derived in Appendix \ref{realLemma}. When we encounter expressions that can be simplified by using these results, we shall usually refer to the relevant formulae in the Appendix and go on with the recursion. We are now ready to proceed. 

We denote the two terms that appear in the simplified Zhu recursion in \eqref{originalZhu} as $I$ and $II$ respectively. We begin with the second term:
\begin{equation}
    II=(n-1)\int d\mu_{1n} \sum_{m_1\ge0}{\cal P}_{m_1+1}\Big(\frac{z_2}{z_1}\Big) \langle (W[m_1]W)(z_2)\ldots W(z_n)\rangle
\end{equation}
Given the contour prescription, we have to do the integrals from $z_1$ to $z_n$ in that order. 
If we consider the integral over $z_1$, we use the fact that 
\be 
\int \frac{dz_1}{z_1} {\cal P}_{m+1}\Big( \frac{z_j}{z_1}\Big) = 0~, \quad m >1~,
\ee 
to restrict the sum to $m=0, 1$. Consider the $m_1=0$ term first. 
The resulting integrated correlator vanishes:
\be 
\int d\mu_{2n} {\cal P}_1\Big(\frac{z_2}{z_1}\Big) \langle (W[0]W)(z_2)\ldots W(z_n)\rangle =0~.
\label{W0WIC}
\ee 
This is courtesy of the Lemma 2 we have used previously; one of the operators is of the form $W[0]X$, with $X=W$ in this case, and all the other operators are identical to $W$.

The remaining term, which we denote $II^{m_1=1}$ is of the form that we are keen to end up with, in which each of the $W$s in the original $n$-point function is replaced by $(WW)_2 \equiv  (2\pi i)^2\, W[1]W$: 
\begin{equation}
    \begin{aligned}
        II &=(n-1)\int d\mu_{1n} {\cal P}_{2}\Big(\frac{z_2}{z_1}\Big) \langle (W[1]W)(z_2)\ldots W(z_n)\rangle\\
        &=(n-1)\int d\mu_{2n}  \langle A^{(2)}(z_2)\ldots W(z_n)\rangle~.
        \label{IIm1=1}
    \end{aligned}
\end{equation}
In the second equality, we have carried out the integral over $z_1$ and introduced some notation. Here and in what follows we will make use of the notation
\be 
 A^{(m)} =(2\pi i)^{2(m-1)}\, (((W[1]W)[1]W)\ldots [1])W 
\ee 
where $m$ is the number of $W$s in this nested product. So, we have $A^{(1)}\equiv W$, $A^{(2)} = (2\pi i)^2\, W[1]W$ and so on. 

We now return to the first term, which has the zero mode insertion in \eqref{originalZhu}, and once more  perform the Zhu recursion, this time using \eqref{zeromodeZhu}:
\begin{equation}
    \begin{aligned}
        I=& \int d\mu_{1n} \langle W_0\, W(z_2)\ldots W(z_n)\rangle\\
        =& 
        \sum_{j=3}^n\int d\mu_{1n} \sum_{m\ge0}{\cal P}_{m+1}\Big(\frac{z_j}{z_2}\Big)\langle W_0\,  W(z_3)\ldots (W[m]W)(z_j)\ldots W(z_n)\rangle\\
        &-(n-2)\int d\mu_{1n} \sum_{m\ge0}g^{1}_{m_2+1}\Big(\frac{z_3}{z_2}\Big)\langle ((W[0]W)[m]W)(z_3)\ldots W(z_n)\rangle + \cO(\tau^2)\\
    \end{aligned}
    \label{termI}
\end{equation}
We have omitted the double zero mode insertion, since we are working only at $\cO(\tau)$, and in the second term, we have again used Lemma 2 to push the composite operator all the way to the left of the correlator. 

Consider the zero-mode term and let us focus just on the $j=3$ term; the $m$ variable is constrained to be $0$ or $1$ on account of the $z_2$-integral. Let us consider the $m=0$ term; to simplify this term we use the operator identity (we use the variables $z=e^{2\pi i u}$; see equation \eqref{AOBflip}):
\begin{equation}
    \begin{aligned}
      \langle\ldots  ( W[0] W)(z) \ldots \rangle  &= \frac{1}{2}\partial_{u}  \langle\ldots (W[1]W)(z)\ldots \rangle + \text{Higher derivatives}
    \end{aligned}
    \label{W0Wasderivative}
\end{equation} 
One can therefore rewrite the integrated correlator with the zero mode insertion and with $m=0$ as 
\be
\begin{aligned}
  I_{1, j=3}^{m=0}= &\frac{1}{2}\int d\mu_{1n} {\cal P}_{1}\Big(\frac{z_3}{z_2}\Big)\partial_{u_3}\langle W_0 (W[1]W)(z_3)\ldots W(z_n)\rangle~.
\end{aligned}
\ee 
We integrate by parts, and use the relation \cite{zhu:1996}
\be 
\partial_u {\cal P}_1(z) = {\cal P}_2(z)~, 
\ee  
to get 
\be
\begin{aligned}
  I_{1, j=3}^{m_2=0}= &-\frac{1}{2}  \int d\mu_{1n}{\cal P}_{2}\Big(\frac{z_3}{z_2}\Big)\langle W_0\, (W[1]W)(u_3)\ldots W(z_n)\rangle\\
  &+\frac{1}{2} (2\pi i) \int d\mu_{2n}\Big[{\cal P}_{1}\Big(\frac{q}{z_3}\Big)\langle W_0\, (W[1]W)(q)\ldots W(z_n)\rangle \\
  &\hspace{5cm}-{\cal P}_{1}\Big(\frac{1}{z_3}\Big)\langle W_0\, (W[1]W)(1)\ldots W(z_n)\rangle\Big]~.
\end{aligned}
\ee 
We use the relation \cite{zhu:1996}
\be 
{\cal P}_1(q z) - {\cal P}_1(z) = 2\pi i~,
\ee 
and obtain 
\be
\begin{aligned}
 I_{1, j=3}^{m_2=0}= &-\frac{1}{2}  \int d\mu_{1n}{\cal P}_{2}\Big(\frac{z_3}{z_2}\Big)\langle W_0\, (W[1]W)(u_3)\ldots W(z_n)\rangle\\
  &+\frac{1}{2} (2\pi i) \int d\mu_{2n}{\cal P}_{1}\Big(\frac{1}{z_3}\Big)\Big[\langle W_0\, (W[1]W)(q)\ldots W(z_n)\rangle \\
  &\hspace{5cm}-\langle W_0\, (W[1]W)(1)\ldots W(z_n)\rangle\Big] + \cO(\tau^2)~.
\end{aligned}
\ee
To simplify the expression within the brackets, one needs to study the periodicity properties of torus correlators involving zero mode insertions. This is analyzed in detail in Appendix \ref{realLemma}, and we have 
\begin{multline}
    \langle W_0\, (W[1]W)(qz)\ldots W(z_n)\rangle -\langle W_0\, (W[1]W)(z)\ldots W(z_n)\rangle \\=-(2\pi i)\langle (W[0](W[1]W))(z)\ldots W(z_n)\rangle     
\end{multline}
The resulting integrated correlator vanishes due to Lemma 2, because the correlator has one operator of the form $W[0]X$, with all other insertions being $W(z_i)$. So we get
\be
\begin{aligned}
 I_{1,j=3}^{m_2=0}= &-\frac12  \int d\mu_{1n}{\cal P}_{2}\Big(\frac{z_3}{z_2}\Big)\langle W_0\, (W[1]W)(z_3)\ldots W(z_n)\rangle+\cO(\tau^2) \\
 = &-\frac12  \int d\mu_{2n}\langle W_0\,  A^{(2)}(z_3)\ldots W(z_n)\rangle+\cO(\tau^2)~.
\end{aligned} 
\ee
In the second equality, we have performed the integral over $z_1, z_2$-integrals, and then reintroduced a free integral over $z_2$.  
Thus, we see that the $m_2=0$ term turns out to be $(-\frac12)$ times the $m_2=1$ term. This analysis can be repeated for each $j$, resulting in 
\begin{equation}
    \begin{aligned}
        I=& \int d\mu_{1n} \langle W_0\,  W(z_2)\ldots W(z_n)\rangle\\
        =& 
         \frac{1}{2}\sum_{j=3}^n\int d\mu_{2n} \langle W_0\,  W(z_3)\ldots A^{(2)}(z_j)\ldots W(z_n)\rangle\\
        &-(n-2)\int d\mu_{1n} \sum_{m\ge0}g^{1}_{m+1}\Big(\frac{z_3}{z_2}\Big)\langle ((W[0]W)[m]W)(z_3)\ldots W(z_n)\rangle + \cO(\tau^2)\\
    \end{aligned}
    \label{Isimplified}
\end{equation}
To further simplify the terms in \eqref{Isimplified}, we substitute for the integral of the function $g^1_{m+1}(z)$ in terms of the Weierstrass functions, by making use of the  identity:
\begin{equation}
    \begin{aligned}
        \int d\mu_{12} g^{1}_{m+1}\Big(\frac{z_3}{z_2}\Big)
        &= \int_{1}^{q}d\mu_{12}\mathcal{P}_{1}\Big(\frac{z_2}{z_1}\Big)\mathcal{P}_{m+1}\Big(\frac{z_3}{z_2}\Big)-\frac{ (2\pi i)^{2}}{2}\int_{1}^{q}\frac{dz_2}{z_2}\mathcal{P}_{2}\Big(\frac{z_3}{z_2}\Big)\delta_{m,2}\\
        &-\left[\frac12(2\pi i)^4 +(2\pi i)^2\int_{1}^{q}\frac{dz_2}{z_2}\mathcal{P}_{1}\Big(\frac{z_3}{z_2}\Big)\right]\delta_{m,1}
    \end{aligned}
\end{equation}
We refer to Appendix \ref{listofintegrals} for a derivation of this identity. 
Let us consider the constant term that contributes for $m=1$; the integrated correlator is of the form
\be 
\int d\mu_{3n}\langle ((W[0]W)[1]W)(z_3)\ldots W(z_n)\rangle ~.
\ee 
Using the relation (see equation \eqref{W0W1})
\be 
\label{01is10}
(W[0]W)[1]W = -\frac12 (W[1]W)[0]W~,
\ee 
one sees that the integrated correlator takes the form
\be 
\int d\mu_{3n}\langle (A^{(2)}[0]W)(z_3)\ldots W(z_n)\rangle  = 0~,
\ee 
which vanishes on account of  Lemma 3 derived in the Appendix (see \eqref{lemmaXOW}). Thus we omit this term and proceed to write the  second term in $I$, involving the $g_{m+1}^1$-function as
\begin{equation}
    \begin{aligned}
        I_2=&-(n-2)\int d\mu_{1n} \sum_{m_2\ge0}\mathcal{P}_{1}\Big(\frac{z_2}{z_1}\Big)\mathcal{P}_{m_2+1}\Big(\frac{z_3}{z_2}\Big)\langle ((W[0]W)[m_2]W)(z_3)\ldots W(z_n)\rangle\\
        &+(n-2)(2\pi i)^2\int d\mu_{2n} \mathcal{P}_{1}\Big(\frac{z_3}{z_2}\Big)\langle ((W[0]W)[1]W)(z_3)\ldots W(z_n)\rangle\\
        &+\frac{(n-2)}{2}(2\pi i)^2\int d\mu_{2n} \mathcal{P}_{2}\Big(\frac{z_3}{z_2}\Big)\langle ((W[0]W)[2]W)(z_3)\ldots W(z_n)\rangle\\
    \end{aligned}
\end{equation}  
We once again use the relation \eqref{01is10} and also $(W[0]W)[2] = -(W[1]W)[1]$ (see \eqref{W0W2}) to obtain:
\begin{equation}
    \begin{aligned}
        I
        =& 
         \frac{1}{2}\sum_{j=3}^n\int d\mu_{2n} \langle W_0\,  W(z_3)\ldots A^{(2)}(z_j)\ldots W(z_n)\rangle\\
        &-(n-2)\int d\mu_{1n} \sum_{m_2\ge0}\mathcal{P}_{1}\Big(\frac{z_2}{z_1}\Big)\mathcal{P}_{m_2+1}\Big(\frac{z_3}{z_2}\Big)\langle ((W[0]W)[m_2]W)(z_3)\ldots W(z_n)\rangle\\
        &-\frac{(n-2)}{2}(2\pi i)^2\int d\mu_{2n} \sum_{m_2\ge0}\mathcal{P}_{m_2+1}\Big(\frac{z_3}{z_2}\Big)\langle ((W[1]W)[m_2]W)(z_3)\ldots W(z_n)\rangle\\
    \end{aligned}
    \label{Iagain}
\end{equation}
Now we use the fact that the second term in the RHS of \eqref{Iagain} can be written as the correlator\footnote{Since we know that  $\langle  W[0]W(z) \ldots \rangle $ is a total derivative, there is no contribution from the zero mode term, we only obtain one term on doing the Zhu recursion.}
\be 
\begin{aligned}
    \int d\mu_{1n} \mathcal{P}_{1}\Big(\frac{z_2}{z_1}\Big)\langle (W[0]W)(z_2) W(z_3)\ldots W(z_n)\rangle ~.
\end{aligned}
\ee 
This vanishes once again as discussed previously. 
Thus the integrated correlator can be written in the form 
\be 
\begin{aligned}
   \frac12 \int d\mu_{12} \mathcal{P}_{1}\Big(\frac{z_2}{z_1}\Big)\int d\mu_{3n}\langle \partial_{u_2}A^{(2)}(z_2) W(z_3)\ldots W(z_n)\rangle ~.
\end{aligned}
\ee 
Such an integrated correlator (with one derivative insertion) vanishes identically on account of Lemma 1 derived in Appendix \ref{realLemma}. Thus, we finally get 
\begin{equation}
    \begin{aligned}
        I =& 
         \frac{1}{2}\sum_{j=3}^n\int d\mu_{2n} \langle W_0\,  W(z_3)\ldots A^{(2)}(z_j)\ldots W(z_n)\rangle\\
        &-\frac{(n-2)}{2}\int d\mu_{2n} \sum_{m_2\ge0}\mathcal{P}_{m_2+1}\Big(\frac{z_3}{z_2}\Big)\langle (A^{(2)}[m_2]W)(z_3)\ldots W(z_n)\rangle~.
    \end{aligned}
    \label{Iintermediate}
\end{equation}
We collect the terms in \eqref{IIm1=1} and \eqref{Iintermediate} and obtain an intermediate result for the integrated $n$-point correlator:
\be
\begin{aligned}
    f_n
        =& 
         \frac{1}{2}\sum_{j=3}^n\int d\mu_{2n} \langle W_0\,  W(z_3)\ldots A^{(2)}(z_j)\ldots W(z_n)\rangle\\
        &-\frac{(n-2)}{2}\int d\mu_{2n} \sum_{m_2\ge0}\mathcal{P}_{m_2+1}\Big(\frac{z_3}{z_2}\Big)\langle (A^{(2)}[m_2]W)(z_3)\ldots W(z_n)\rangle\\
        &+(n-1) \int d\mu_{2n}  \langle A^{(2)}(z_2)W(z_3)\ldots W(z_n)\rangle + \cO(\tau^2)
\end{aligned}
\ee 
While we have managed to obtain lower point integrated correlators on the right hand side, this is not the recursion we are after; what we are aiming for is to have only field insertions on the right hand side, and not correlators with zero mode insertions. To address this, we begin with an auxiliary integrated correlator below and do the Zhu recursion once:
\be 
\begin{aligned}
    \int d\mu_{2n} \langle &W(z_2)\ldots A^{(2)}(z_j)\ldots W(z_n)\rangle =  \int d\mu_{2n}\langle W_0\, W(z_3) \ldots A^{(2)}(z_j)\ldots W(z_n)\rangle \\
    &+  \int d\mu_{2n}\sum_{m} {\cal P}_{m+1}\Big(\frac{z_{j}}{z_2}\Big) \langle W(z_3) \ldots (W[m]A^{(2)})(z_j) \ldots W(z_n) \rangle \\
    &+\sum_{k\ne j}\int d\mu_{2n}\sum_{m} {\cal P}_{m+1}\Big(\frac{z_{k}}{z_2}\Big) \langle W(z_3)\ldots A^{(2)}(z_j)\ldots (W[m]W)(z_k) \ldots W(z_n) \rangle \\
\end{aligned}
\ee 
In the second term, the composite operator can again be moved onto the leftmost position in the correlator without cost since the terms one get from commuting this operator through is of the form that vanishes by the Lemma discussed previously in \eqref{WOXlemmaintext}. 
In the third term, the integral over $z_2$ imposes $m = 0,1$. The $m=0$ term once again vanishes, since the vertex operator associated to the $W[0]W$-operator is a total derivative (see \eqref{W0Wasderivative}), and so only the $m=1$ term survives. Thus, by rearranging the terms, we obtain an expression for the integrated correlator with the zero mode insertion:
\be 
\begin{aligned}
   \int d\mu_{2n}\langle W_0\, W(z_3) \ldots &A^{(2)}(z_j)\ldots W(z_n)\rangle  =   \int d\mu_{2n} \langle W(z_2)\ldots A^{(2)}(z_j)\ldots W(z_n)\rangle\\
    &-  \int d\mu_{2n}\sum_{m} {\cal P}_{m+1}\Big(\frac{z_{3}}{z_2}\Big) \langle (W[m]A^{(2)})(z_3) W(z_4)\ldots W(z_n) \rangle \\
    &-\sum_{k\ne j}\int d\mu_{3n}\langle W(z_3)\ldots A^{(2)}(z_j)\ldots A^{(2)}(z_k) \ldots W(z_n) \rangle ~.
\end{aligned}
\ee 
Substituting this into the expression for $f_n$, we find 
\be
\begin{aligned}
     f_n
        =& 
        \Big(\frac{3n}{2}-2\Big)\int d\mu_{2n} \langle A^{(2)}(z_2)W(z_3)\ldots \ldots W(z_n)\rangle\\
        &-\frac{(n-2)}{2}\int d\mu_{2n} \sum_{m\ge0}\mathcal{P}_{m+1}\Big(\frac{z_3}{z_2}\Big)\Big[\langle   (A^{(2)}[m]W)(z_3)W(z_4)\ldots W(z_n)\rangle\\
        &\hspace{5cm} +\langle   (W[m]A^{(2)})(z_3) W(z_4) \ldots W(z_n) \rangle\Big]\\
        &-\frac12 \sum_{k\ne j}\int d\mu_{3n}\langle W(z_3)\ldots A^{(2)}(z_j)\ldots A^{(2)}(z_k) \ldots W(z_n) \rangle + \cO(\tau^2)
\end{aligned}
\ee 
To write the first term, we have commuted the $A^{(2)}$ operator all the way to the left without generating additional terms as usual, using Lemma 2. 
In the terms appearing in the second and third lines, the integral over $z_2$ ensures that $m =0, 1$. When $m=0$, we have  
\be 
\langle  \Big[ (A^{(2)}[0]W)(z_3) +W[0]A^{(2)}(z_3)\Big] W(z_4)\ldots W(z_n)\rangle = \langle   \partial \cO(z_3)W(z_4)\ldots W(z_n) \rangle ~. 
\ee 
The integrated correlator with a single total derivative term vanishes as shown in Lemma 1 of the Appendix \ref{realLemma}.   
Thus, only the $m=1$ terms survive in these two terms; for this case:
\begin{equation*} 
\begin{aligned}
\langle W[1]A^{(2)}(z_3)W(z_4) \ldots W(z_n)\rangle &=\langle  (A^{(2)}[1]W)(z_3)W(z_4) \ldots W(z_n)\rangle + \langle   \partial \tilde \cO(z_j)W(z_4)\ldots W(z_n) \rangle \\
&= \frac{1}{(2\pi i)^2}\langle A^{(3)}(z_3)W(z_4) \ldots W(z_n)\rangle+ \langle   \partial \tilde \cO(z_j)W(z_4)\ldots W(z_n) \rangle~.
\end{aligned}
\end{equation*} 
The integrated correlator involving the total derivative vanishes,  
and performing the integral over $z_2$ we find that the integrated $n$-point correlator can be written as 
\be 
\begin{aligned}
  f_n =& 
        \Big(\frac{3n}{2}-2\Big)\int d\mu_{2n} \langle A^{(2)}(z_2)W(z_3)\ldots \ldots W(z_n)\rangle\\
   &-(n-2)\int d\mu_{3n} \langle A^{(3)}(z_3)W(z_4)\ldots W(z_n) \rangle \\ 
&-\frac12 \sum_{k\ne j}\int d\mu_{3n}\langle W(z_3)\ldots A^{(2)}(z_j)\ldots A^{(2)}(z_k) \ldots W(z_n) \rangle  + \cO(\tau^2) ~.
\end{aligned}
\label{npointaslowerpoint}
\ee 
This is the main result of the Zhu recursion. We have shown that, up to $\cO(\tau^2)$ terms, the integrated $n$-point correlator can be written in terms of a sum of lower point integrated correlators. 

We now claim an important generalization of this recursive result for the integrated $n$-point correlator:    equation 
\eqref{npointaslowerpoint} holds for an arbitrary local operator $W(z)$, and is not restricted to operators with a fixed scaling dimension. We prove this result in detail in Appendix \ref{generalizationofZhuresult} by showing that every step used in the recursion of the integrated $n$-point correlator also holds for an arbitrary local field.

\section{Proof of the Conjecture}
\label{provingconjecture}

The next step in our analysis is to rewrite the recursion in terms of a variational derivative. This will allow us to write the recursion relation for the integrated correlator in a compact form, which in turn will allow us to prove the conjecture using induction. 

\subsection{The Variational Derivative}

Let us take a step back and start with an integrated correlator of $n$ arbitrary fields $X_i(z_i)$. Then by performing Zhu recursion all the way to the end, the $\cO(\tau)$ coefficient will take the schematic form: 
      \begin{equation}\label{E2}
  \begin{gathered}
        \int d\mu_{1n}\langle X_1(z_1)\dots X_{n}(z_{n}) \rangle = \sum_I \tau \, g_I \langle (((X_{k_1}[\ell_1]X_{k_2})[\ell_2]...X_{k_{n-1}})[\ell_{n-1}]X_{k_n})_0\rangle + \mathcal{O}(\tau^2)
        \end{gathered}
    \end{equation}
Here $I$ is a multi-index $(k_1,\dots, k_{n},\ell_1,\dots, \ell_{n-1})$ such that $k_j$ is a permutation of $1,\dots,n$ and $\ell_k \geq 0$. Importantly, the numbers $g_I$ depend only on the multi-index $I$ and are independent of the local fields one might choose to be $X_1(z_1),\dots, X_{n}(z_{n})$. That \eqref{E2} is true follows simply by noting that the $g_I$ come about through the integrations over the generalised Weierstrass functions brought about by evaluating the $n$ point correlator using the Zhu recursion. In the Zhu formula, these Weierstrass functions can be seen to depend only on the multi-index $I$.  The application of \eqref{E2} to the integrated $n$-point correlator leads to 
\begin{equation}
\begin{gathered}
     \int d\mu_{1n} \langle W(z_1)\dots W(z_n) \rangle = \sum_J  \tau \, g_J \, \langle (((W[i_1]W)[i_2]\dots  [i_{n-1}]W)(z_{n})\rangle + \mathcal{O}(\tau^2)
    \end{gathered}
        \label{nZhu}
\end{equation}
where the $g_J$'s here are the same as those in \eqref{E2}, because the $g_J$ are independent of the choice of local fields involved in the $n$-point correlator.

We are now in a position to formally  define the notion of a variational derivative. Consider a formal expression of the form:
\begin{equation} \label{form}
    (W[i_i](W[i_1]\dots W[i_n]W))[i_{n+1}]...(\dots W)[i_{n+m}]W)(z)~.
\end{equation}
The operator 
    \be 
    \delta_W := (2\pi i)^2\, W[1]W\frac{\p}{\p W}
    \ee
acts on this formal expression as follows: (i) we first replace each local field $W$ in this expression by the linear combination $W + \epsilon \,  (2\pi i)^2\, W[1]W$, and then (ii) collect  linear order terms in $\epsilon$. More explicitly, 
\begin{equation}
\begin{aligned}
       \delta_{W} &\cdot \left((W[i_i](W[i_1]\dots W[i_n]W))[i_{n+1}]...(\dots W)[i_{n+m}]W)(z)\right) \\ &\hspace{2cm}:=  \lim_{\epsilon \to 0} \frac{1}{\epsilon} \Big[ \left((X_\epsilon[i_i](X_\epsilon[i_1]\dots X_\epsilon[i_n]X_\epsilon))[i_{n+1}]...(\dots X_\epsilon)[i_{n+m}]X_\epsilon)(z))\right) \\ &\hspace{2.2cm}- \left((W[i_i](W[i_1]\dots W[i_n]W))[i_{n+1}]...(\dots W)[i_{n+m}]W)(z)\right) \Big]
\end{aligned}
\end{equation}
where $X_\epsilon(z) := W(z) + \epsilon \, (2\pi i)^2\, W[1]W(z)$. The $\delta_{W}$ operation also has the property of being distributive over formal sums and $\mathbb C$-linear. Crucially, the object on the RHS is to be interpreted as an element of the OPE algebra of $W$ (as we do throughout the main text).  What this means is that one can now act with the variational derivative $\delta_W$ on the $\cO(\tau)$ terms of the integrated correlators. The takeaway from this section is that the more general validity of the recursive relation for integrated correlators (now applicable to local operators) allows us to formalize the replacement operation discussed heuristically in Section \ref{prelims}.

\subsection{The Two-term Recursion}

With this in hand, let us return to the recursion relation that we obtained for the integrated $n$-point correlator in \eqref{npointaslowerpoint} and rewrite it in the following manner:
\begin{align}
   f_n  =& 
        \Big(\frac{3n}{2}-2\Big)\frac{(2\pi i)^2}{n-1}\sum_{j=1}^{n-1} \int d\mu_{1,n-1} \langle W(z_1)\ldots (W[1]W)(z_j)\ldots W(z_{n-1})\rangle\cr
   &\hspace{.5cm}-(2\pi i)^4\sum_{j=1}^{n-2}\int d\mu_{1,n-2} \langle W(z_1)\ldots ((W[1]W)[1]W)(z_j)\ldots W(z_{n-2}) \rangle \cr
&-(2\pi i)^4\frac12 \sum_{\substack{j,k=1 \\ j \neq k}}^{n-2}\int d\mu_{1,n-2}\langle W(z_1)\ldots (W[1]W)(z_j)\ldots (W[1]W)(z_k) \ldots W(z_{n-2}) \rangle +\cO(\tau^2) ~.\nonumber
\end{align}
We have used the Lemma in \eqref{WOXlemmaintext} to write the correlators in a symmetric fashion, and also used the definition of $A^{(k)}(z)$ to rewrite it in terms of the nested $W[1]W$ operators. 

In the first term on the right hand side of this equation, we have an  $(n-1)$-point function in which $W$ has been replaced by $W[1]W$ at each position. Similarly, the remaining two terms involve $(n-2)$-point functions in which there has been a successive replacement of $W$ by $W[1]W$.
Thus one can check that the recursive relation between integrated correlators can be written using the variational derivative $\delta_W$ in a rather compact form: 
\be 
\begin{aligned}\label{ourresult}
f_n =&  \Big(\frac{3n}{2}-2\Big)\frac{1}{n-1} \delta_W\cdot f_{n-1}
-\frac{1}{2}\delta_W\cdot\delta_W\cdot f_{n-2} + \mathcal{O}(\tau^2)~.
\end{aligned}
\ee  
The boundary case is easily evaluated directly:
\begin{equation}
    f_2 = \delta_W \cdot f_1 + \mathcal{O}(\tau^2)~.
\end{equation}
From the form of the recursion and the boundary case, it is straightforward to recurse the higher $f_n$ from the lower ones, and to write $f_n$ as a higher order variational derivative acting solely on $f_1$; in fact it is possible to write a simple ansatz that solves the recursion: 
\begin{equation}
    f_n = \alpha(n)~\delta_W^{n-1}\cdot f_1 + \mathcal{O}(\tau^2)
\end{equation}
Computing the first few $f_n$, we find 
\begin{equation}
\begin{aligned}
    \alpha(3) &= \frac{3}{4} ~, \quad 
    \alpha(4) &= \frac{1}{2}~, \quad 
    \alpha(5) &= \frac{5}{16} 
   ~.
\end{aligned}
\end{equation}
The pattern is clear and one can proceed by induction to prove that 
\be \alpha(n) = \frac{n}{2^{n-1}}~.
\ee 
Let us assume that the ansatz holds for $f_n$. Then for $f_{n+1}$ we find that 
\be 
\begin{aligned}\label{ourresult1}
f_{n+1} =&  \Big(\frac32(n+1)-2\Big)
\frac{1}{n}\delta_W\cdot f_n
-\frac{1}{2}\delta_W\cdot \delta_W\cdot f_{n-1} + \mathcal{O}(\tau^2) \\  =&  \Big(\frac32(n+1)-2\Big)\frac{1}{2^{n-1}}\delta_W^n f_1
-\frac{1}{2}\frac{n-1}{2^{n-2}} \delta_W^n f_1 + \mathcal{O}(\tau^2) \\ =& \frac{n+1}{2^n}\delta_W^n f_1 + \mathcal{O}(\tau^2)~,
\end{aligned}
\ee  
completing the inductive step. 

Let us now recall equation \eqref{ordertauintegcorr}, which resulted from  the  proposal for the S-transformation of the GGE. 
The $\cO(\tau)$ term of the integrated $(n+1)$-point correlator determines the $(n+1)$st term in the asymptotic expansion of the operator in the modular transformation of the generalized partition function:
\be 
f_{n+1} = \frac{n+1}{2^{n}}\langle [{ {W}}^{n+1}] \rangle\, (2\pi i\tau) + \cO(\tau^2) ~.
\ee 
Thus, a recursion for $f_{n}$ implies a recursion for the composite operator $[W^n]$, and we obtain that the composite operator $[W^n]$ satisfies the recursion:
\be 
\langle [W^{n+1}] \rangle = 
 \delta_W^{n}\cdot \langle W\rangle ~,
\ee 
thereby proving the conjecture in \eqref{Wnrecursionv3}. 

\section{Discussion}
\label{conclusions}

In this concluding section, we shall begin by summarizing the various steps that led us to  the iterative relation that determines the modular transformation of the generalized partition function. We shall then show that our results are applicable to arbitrary chiral deformations of a conformal field theory. We discuss two simple applications of our recursive formula, making contact with the previous results in the literature, and provide an outlook for future work.  

\subsection{Summary}

We began with the generalized partition function, which is the torus correlator given by: 
\be 
\label{bootGGE}
\langle e^{\alpha W_0} \rangle ~.
\ee 
Here $W_0$ is the zero mode of an arbitrary (integer) spin-$w$ current $W(z)$. 
First of all, we have restricted ourselves to the asymptotic expansion of this generalized partition function in $\alpha$, and focused on the S-transform of the zero-mode correlator
\be 
\langle W_0^n \rangle ~.
\ee 
The S-transform of the zero mode correlator is obtained by computing the integrated $n$-point correlator of $W(z_i)$, where the integrals are  over the $B$-cycle of the torus. We showed on general grounds that the S-transform of the integrated $n$-point correlator has a particular form: an expansion in $\tau$, with coefficients given by quasimodular forms of a certain degree (at each order in the $\tau$-expansion). 

At this point we made an important assumption\footnote{We provide arguments for this assumption in section \ref{sec:outlook}}: that the result of the modular transformation could be written as the exponential of the zero mode of an operator, given by 
\be 
\langle e^{\alpha  {\cal W}_0}\rangle ~.
\label{RHSofStransform}
\ee 
Here ${\cal W}$ is a local field, which has an asymptotic expansion in $\alpha$, given by 
\be 
{\cal W} = \sum_{n=0}^{\infty}\frac{1}{n!} \left(\frac{\alpha}{4\pi i\tau}\right)^n\, [W^{n+1}]~.
\label{calWdefn2}
\ee 
Here, the composite operators $[W^{n+1}]$ are undetermined at this point. 
Given the general form taken by the S-transform of $\langle W_0^n\rangle$, and the assumption regarding the action of the modular transformation in \eqref{RHSofStransform}, we showed that the $\cO(\tau)$-coefficient of the integrated $n$-point correlator is proportional to $[W^n]$. Thus, our focus for the majority of this work has been to isolate the $\cO(\tau)$ coefficient of the integrated $n$-point correlator and to study its recursive properties. 

The next step was to use the Zhu recursion relation to write the integrated $n$-point correlator as a sum over $(n-1)$ and $(n-2)-$point integrated correlators, up to terms that are $\cO(\tau^2)$. This recursion in \eqref{npointaslowerpoint} is the main technical result of our work, and one that is obtained after going through a maze of non-trivial identities  involving integrals, correlators and integrated correlators. Importantly, we showed that the derived recursion is, in fact, valid for $W(z)$ being an arbitrary local field and not just for operators with fixed conformal weight. This led naturally to the formal definition of a variational derivative $\delta_W$,  that in turn allowed us to rewrite the recursive relation between integrated correlators as a compact 2-term recursion in equation \eqref{ourresult}. The solution to the recursion directly led us to the fact that 
\be 
\langle [W^{n+1}]\rangle  =\delta_W\cdot \langle [W^{n}] \rangle ~,
\label{Wnrecursion}
\ee
which proves the conjecture.

We have proven that the one point function of the $n$th term in the expansion \eqref{calWdefn2} of ${\cal W}$ satisfies the expected recursion. 
This does not mean that we have uniquely identified operators $Q_n$ which satisfy
\begin{align}
\langle Q_n \rangle = \langle [W^n]\rangle
\;,
\label{eq:Qn}
\end{align}
nor uniquely identified
an operator ${\cal Q}$ such that 
\begin{align}
    \langle e^{\frac{\alpha}{\tau^w}\, W_0} \rangle_{-\frac{1}{\tau}}   
    = \langle e^{ {\cal Q}} \rangle_{\tau}\;.
    \label{eq:allQ}
\end{align}
If $Q_n$ satisfies \eqref{eq:Qn}, then so does $Q'_n = Q_n + D_n$ where $D_n$ is any operator with zero trace, 
and 
if ${\cal Q}$ satisfies \eqref{eq:allQ} , then so does $ {\cal Q}'$ defined by
\begin{align}
    e^{{\cal Q}'} = e^{ D} e^{{\cal Q}} e^{-D}\,
    \label{eq:simil}
\end{align}
for any operator $D$ that commutes with $L_0$. Using the Baker-Campbell-Hausdorff formula, ${\cal Q}$ and ${\cal Q}'$ are related by
\begin{align}
     {\cal Q}' = e^{\text{ad}_{D}}({\cal Q})
    = {\cal Q} + [  D,  {\cal Q}] + \tfrac 12 [  D,[  D,  {\cal Q}]] + \ldots
    \;.
    \label{eq:defdef}
\end{align}
As the extra terms are all commutators, they do not contribute to the trace of $  {\cal Q}$.
As a simple example, if $  {\cal Q} = \alpha {\cal W}_0$ and $  D = a \Lambda_0$, then this can be understood order by order in $\alpha$ as the changes
\begin{align}
\begin{split}
    [W] = W & \mapsto 
    [W]_a = W + a [\Lambda_0,W] + \tfrac{a^2}{2} [\Lambda_0,[\Lambda_0,W]] + \ldots\;,\\
    [W^2] = (WW)_2 &\mapsto 
    [W^2]_a = 
    [W^2] + a [\Lambda_0,[W^2]] + \tfrac{a^2}{2} [\Lambda_0,[\Lambda_0,[W^2]]] + \ldots\;,
    \end{split}
    \label{eq:w-deformed}
\end{align}
The charges formally derived from these fields will themselves be commutators, and hence, all clearly have zero trace. The new fields  would appear at first sight from \eqref{eq:w-deformed} to
obey the same recursion relation as the one-point functions of the original fields,
\begin{align}
    [W^{n+1}]_a = \delta_W \cdot [W^n]_a
    \;,
\end{align}
except that $\delta_W$ is only defined on nested products of $W$, not commutators with further generators and so it is not clear that it can be defined on the deformed charges that appear in  \eqref{eq:w-deformed}. It is also not clear if the series defined by \eqref{eq:defdef} is convergent, but it does demonstrate one consistent way in which commutator terms can be added to our solution to the recursion relation without changing the one-point functions or the exponentiation property.

It is a fact that any finite dimensional matrix with zero trace can be expressed as a commutator (see eg \cite{AM}); we also know that
the commutator of two zero modes is equal to the zero mode of a local field, namely the zero mode of the coefficient of the first order pole in the OPE of the two original local fields,
\begin{align}
    [A_0,B_0] = [(AB)_1 ]_0\;,
\end{align}
and so it is plausible that any local field $C$ for which the trace of $C_0$ vanishes can be written as a sum of such terms, so that
\begin{align}
    C_0 = \sum_i [A^i_0,B^i_0]
    \;.
\end{align}
To return to \eqref{Wnrecursion} - we have found a solution to this recursion relation for the one-point functions of the fields, but we cannot expect it to be unique. Just based on the one-point functions of the fields $\langle[W^n]\rangle$, we could add any field with vanishing trace to $[W^{n+1}]$, but the requirement that this does not affect the one-point function of the exponential is very strong and could well limit the ambiguity to the similarity transformation \eqref{eq:simil}.

\subsection{Chiral Deformations of a Conformal Field Theory}

Let us now discuss the generalization of our results to the case in which $W_0$ is a linear combination of zero modes of operators:
\be 
W(z) = \sum_{\ell} \alpha_{\ell} I^{\ell}(z) ~,
\ee 
where, on the right hand side, we have  quasiprimaries $I^{\ell}(z)$. This would correspond to a deformation of the conformal field theory by a local (holomorphic) field. 
We begin with the generating function: 
\be 
\label{gengge}
\langle  e^{\lambda\, W_0} \rangle ~.
\ee 
Expanding out the generating function, we have 
\begin{align}
\langle  e^{\lambda\, W_0} \rangle=\sum_{n=0}^{\infty} \frac{\lambda^n}{n!} \langle W_0^n\rangle &= \sum_{n=0}^{\infty} \frac{\lambda^n}{n!} \sum_{i_1\ldots i_n} \alpha_{i_1}\ldots \alpha_{i_n} \langle I_0^{i_1}\ldots I_0^{i_n} \rangle_S ~,
\end{align}
where the subscript S on the correlator indicates symmetrization. This ensures that the correlator will be a quasimodular form \cite{Dijkgraaf:1996iy}. The zero mode correlator can be written as the integrated correlator over the $A$-cycles of the torus:
\be 
\langle I_0^{i_1}\ldots I_0^{i_n} \rangle_S = \int_A d\mu_{1n} \langle I^{i_1}(z_1)\ldots I^{i_n}(z_n) \rangle_S ~.
\ee 
We now consider the S-transformation, that acts on the fugacities in the following manner\footnote{We consider $\lambda$ to be a dummy variable that does not transform under modular transformation.}: 
\be 
S:~\tau \rightarrow -\frac{1}{\tau}~,\quad \alpha_{\ell} \rightarrow \frac{\alpha_{\ell}}{\tau^{w_{\ell}}}~. 
\ee 
Under this action, since each of the $I^{i_{\ell}}$ is a quasiprimary of weight $w_{\ell}$, the integrated correlator transforms homogeneously; following through the change of variables (see the discussion below equation \eqref{correlatortransform}), we obtain a sum of integrated correlators, where we once again end up with integrals over the $B$-cycles of the torus:
\begin{align}
S\left[\langle W_0^n\rangle\right] &= \frac{1}{(2\pi i \tau)^n}\int_B  d\mu_{1n}\sum_{i_1\ldots i_n} \alpha_{i_1}\ldots \alpha_{i_n} \langle I^{i_1}(z_1)\ldots I^{i_n}(z_n) \rangle_S \cr 
&= \frac{1}{(2\pi i \tau)^n}\int_B  d\mu_{1n} \langle W(z_1) \ldots W(z_n) \rangle~.
\end{align} 
The transformation of the fugacities under the modular S-transformation is crucial in order to have the uniform $\tau$-scaling in the first line. In the second equality, we have simply resummed the correlator back into a correlator involving the local field $W(z)$. 
Now we focus on the $\cO(\tau)$ contribution, and we can recurse this correlator exactly as before. We have already shown that the Zhu recursion can be applied to a local field. Thus our previous result, in equation \eqref{npointaslowerpoint},   continues to hold and we propose that the modular transformation of the generalized partition function of the deformed conformal field theory is given by 
\be 
\langle e^{\lambda {\cal W}_0} \rangle,
\ee 
where ${\cal W}$ is given by
\be 
{\cal W} = \sum_{n=0}^{\infty} \frac{\lambda^n}{n!} \frac{1}{(4\pi i \tau)^n}\, [W^{n+1}],
\ee  
with the successive terms in the asymptotic expansion obtained by recursion
\be 
\langle [W^{n+1}]\rangle = \delta_W \cdot\langle [W^n]\rangle ~.
\ee 
In other words we claim that our results for the modular transform hold for an arbitrary chiral deformation of the conformal field theory.

\subsection{A \texorpdfstring{$U(1)$}{U(1)} Example}

Let us discuss a simple application of our result. Consider a $U(1)$ current algebra at level $k$, generated by a current $J(z)$ that has the operator product expansion: 
\be 
J(z)J(w) \sim \frac{k/2}{(z-w)^2} + \ldots~. 
\ee
Thus the second order pole is a constant and equal to $\frac{k}{2}$. As a consequence, our recursive formula truncates at the first order,  and we have 
\be 
{\cal W} = W + \frac{k\alpha}{8\pi i\tau}~. 
\ee 
In order to make contact with the literature, let us set $\alpha = 2\pi i \beta$. Then, we find that\footnote{This relation also holds true as an exact equality, while our result only implies the equivalence of the asymptotic expansions.} :
\begin{equation}
    S\left[\langle e^{2\pi i\beta J_0}\rangle_{\tau} \right] \sim e^{  \frac{\pi i k\beta^2}{\tau}} \langle e^{2\pi i \beta J_0}\rangle_\tau~,
\end{equation}
which is the expected transformation law. 

\subsection{Functional Relations for Chiral Deformations}

In the U$(1)$ example, the fact that the coefficient of the second order pole in the OPE of $J$ with itself is a constant means that we avoid  the introduction of new fields into the S-transformed partition function. This allowed for  the modular transformation to be rewritten as a functional relation for the partition function. Given our result for the modular transformation, let us now explore what type of functional relation is obtained for the partition function resulting from a generic chiral deformation of a conformal field theory. 

Denote by $\mathcal{F}$ the space of  holomorphic fields by which we deform the conformal field theory; then consider the minimal subspace ${\cal J}$ of the space of holomorphic fields  such that :   
\begin{equation}
\begin{aligned}
       \mathcal{F} &\subseteq \mathcal J \\ A, B &\in  {\cal J}~ \implies (AB)_2 \in \mathcal J
\end{aligned}
\end{equation}
and let $J^i, i \in \I$ be a basis of $\mathcal J$ with some indexing set $\I$. We consider a general chiral deformation of the conformal field theory by 
\be 
W(z) = \sum_{i\in {\cal I} }\alpha_i J^i(z)~.
\ee 
The S-transform formula implies a functional relation on the generalized partition function:
 \begin{equation}
Z[\tau,\alpha_1,\dots,\alpha_{\ell}] := \Big \langle \exp \left(\sum_{i \in \I} \alpha_i J^i_0\right) \Big\rangle_\tau~.
 \end{equation}
To write the functional relation more explicitly, we need to introduce some notation. We start by defining ``structure constants"  :
 \begin{equation}
       (J^iJ^j)_2 = \sum_{k\in \I} c^{ij}_k J^k\,,
 \end{equation}
 In writing down the functional relation, it is useful to consider rescaled structure constants
 \be 
 \beta_k^{ij} :=c_k^{ij}\frac{\alpha_i\alpha_j}{\alpha_k}~,
\ee
 where the repeated $(i,j)$ indices are not summed over. 
In order to derive the functional relation, what we shall aim to do is to rewrite the modular transformation as an action on the fugacities. For this, one needs to express ${\cal W}$ in terms of the basis elements $J^k$ of the operator algebra and read off the coefficient as the transformed $\alpha_k$. For instance, at the first step, we have:
\be 
[W^2] = \sum_{i,j,k\in \I} \alpha_i \alpha_j c^{ij}_k J^k = \sum_{i,j,k\in \I} \beta^{ij}_k \alpha_k J^k ~. 
\ee 
The operator that picks out the coefficient of $J^k$ can be written in the symmetric form:
\begin{equation}
    \frac{1}{2}\sum_{i,j\in \I}\beta^{ij}_k \alpha_k \frac{\p}{\p \alpha_i} \frac{\p}{\p\alpha_j} \sum_{I_i \in \I} \alpha_{I_1}\alpha_{I_2} ~. 
\end{equation} 
In order to obtain the result after the $n$th iteration, one notes the formal similarity between this operator and the one associated to Wick contractions. Motivated by this we define the  function:
\begin{equation}
\begin{aligned}
    \kappa(1,m) &= \alpha_m \\
    \kappa(n,m) &= \frac{1}{n!}\sum_{I_i \in \I}\left(\sum_{i,j\in \I}  \beta^{ij}_m \alpha_m \frac{\p}{\p \alpha_i} \frac{\p}{\p\alpha_j}\right) \left(\sum_{i,j,k\in \I}\beta^{ij}_k \alpha_k \frac{\p}{\p \alpha_i} \frac{\p}{\p\alpha_j}\right)^{n-2}  \prod_{i=1}^n \alpha_{I_i}, \quad n\geq 2
    \end{aligned}
    \label{kappadefn}
\end{equation}
This function simply accounts for the Wick contractions (involving the second order pole) between the currents involved. Importantly, the derivatives $\frac{\p}{\p \alpha_j}$ do not act on the structure constants $\beta^{ik}_\ell$. While not obvious from equation \eqref{kappadefn}, it is possible to prove that the $\kappa(n,m)$ expressed in terms of $c^{ij}_k$s have a polynomial dependence on the $\alpha_j$'s \footnote{This follows from the fact that $\kappa(n,m)$ is a product of $n-1$ $\beta^{ij}_k$ and that all but one of the subscripts (the uncontracted subscript takes value $m$, but $\kappa(n,m)$ comes with an explicit $\alpha_m$ factor) are contracted, so the expression in terms of $c^{ij}_k$ will involve inverse powers of the $\alpha_j$s.}.

We are now ready to write down the functional relation satisfied by the generalized partition function:
\begin{equation}
    Z\left(-\frac{1}{\tau},\frac{\alpha_1}{\tau^{h_1}},\dots \frac{\alpha_{\ell}}{\tau^{h_{\ell}}}\right) \sim Z\left(\tau, \alpha_1+\sum_{n\geq1} \frac{1}{n!}\frac{ \kappa(n+1,1)}{(4\pi i\tau)^n}, \dots, \alpha_{m} +  \sum_{n\geq 1} \frac{\kappa(n+1,{m})}{n! (4\pi i\tau)^n},\ldots \right)~.
    \label{generalfunctionalrelation}
\end{equation}
The ellipses on the right hand side of the functional relation indicate that, even if one starts with a small but finite set of holomorphic deformations (with most of the $\alpha_m$ being set to zero), the modular transform leads to a much larger set of deformations. The $\alpha_m$ that are turned on depend on the non-vanishing second order pole coefficients in the operator product algebra, which are encoded in the $\kappa(n,m)$. For spin $>2$ deformations, the functional relation does not close on a finite number of fields on the right hand side. Despite the complicated and non-linear action on the fugacities, one can check that acting with the S-transformation twice, the fugacities transform back to themselves up to charge conjugation  ($\alpha_m \rightarrow  (-1)^{h_m}\alpha_m$). 

The functional relation in \eqref{generalfunctionalrelation} is entirely equivalent to the iterative definition of ${\cal W}$. By choosing a basis for the operator algebra and the associated second order pole coefficients in the operator product expansion, we have restated the action of the modular transformation as a non-trivial (and non-linear) transformation of the fugacities associated to the basis elements of the operator algebra.   

\subsubsection{Modular Properties of \texorpdfstring{$W_{1+\infty}$}{W1plusinfinity} Characters}

To illustrate the working of the functional relation, and the action of the modular transform on the fugacities,  let us consider the $W_{1+\infty}$ algebra at $c=1$, whose modular properties were studied in \cite{Dijkgraaf:1996iy}. The algebra is  generated by the bosonic currents 
\be 
J^{\ell} = \frac{1}{\ell+1}(-i\,  \partial\phi)^{\ell+1} \quad\text{mod}\quad \partial~.
\ee 
These satisfy the algebra 
\be 
J^i(z)J^j(w)\sim \ldots (i+j)\, \frac{J^{i+j-1}(w)}{(z-w)^2} + \ldots 
\ee 
We study the modular properties of the partition function
\be 
Z(\tau, \alpha_i) = \left\langle \exp(\sum_{i} \alpha_i J^i) \right\rangle~.
\ee 
Following \cite{Dijkgraaf:1996iy}, we set $\alpha_1=0$. Let us see how the other $\alpha_k$ transform under modular transformation. This involves computing the coefficient of the $J^k$ after modular transformation. The only way to generate $J^2$ as a second order pole is the OPE of $J^1$ with $J^2$. But since $\alpha_1=0$, this contribution is absent and so $\alpha_2$ is not transformed. For $\alpha_3$, the modular transformation involves calculating the coefficient $\kappa(i,3)$. The relevant $\beta^{ij}_k$ for the computation will be :
\begin{align}
    \beta^{ij}_3 =  \frac{4\alpha_i \alpha_j}{\alpha_3}, \ \text{for} \ i+j=4~.
\end{align}
Clearly $(i,j)$ can take values $(1,3), (3,1), (2,2)$. Since $\alpha_1=0$, only the last pair contributes. 
From the definition of the $\kappa(i,j)$ we see that the only non-trivial $\kappa$ is 
\begin{align}
    \kappa(2,3) &= \frac{1}{2!} \sum_{i,j}\beta^{ij}_3 \,\alpha_3\,  \frac{\partial^2}{\partial\alpha_i\partial\alpha_j} (\alpha_2^2)\cr 
    &= \alpha_3\, \beta^{22}_3 = 4\alpha_2^2~.
\end{align}
A similar analysis can be carried out for the higher currents as well, and we find the modular transform: 
\begin{align} 
Z(-\frac{1}{\tau}, \frac{\alpha_2}{\tau^3},\frac{\alpha_3}{\tau^4}, \frac{\alpha_4}{\tau^5}\ldots) \sim Z\left(\tau, \alpha_2, \alpha_3 + \frac{\alpha_2^2}{\pi i\tau}, \alpha _4 -\frac{5 i \alpha _3 \alpha _2}{2 \pi  \tau }-\frac{5 \alpha _2^3}{4 \pi ^2 \tau ^2}, \ldots \right)
\end{align} 
The transformation of the fugacities we have derived agrees\footnote{In the conventions of \cite{Dijkgraaf:1996iy}, the charges are defined with an extra factor of $-(2\pi)$. Taking care of this factor, one finds an exact match between our results and those in \cite{Dijkgraaf:1996iy}.} with the one derived in \cite{Dijkgraaf:1996iy}. 
This earlier study involved an intermediate step, relating perturbations by integrals over the plane to insertions of charges, in the special case that the currents generated a pre-Lie algebra. In this work, we have related the line integrals directly for general currents and found a closed form expression for the modular transformed partition function for generic chiral deformations in equation \eqref{generalfunctionalrelation}. 
Thus our results should be thought of as an extension of the results in \cite{Dijkgraaf:1996iy} to a wider class of algebras. 

\subsection{Outlook}
\label{sec:outlook}

The functional relation discussed in the previous  subsection  highlights the point that even if one had started with a small set of non-zero fugacities, the modular transform would turn on the  fugacities associated to higher spin deformations. 
For the GGEs defined by a set of integrals of motion in involution, this was already observed, that the S-transformation takes one out of the GGE, and involves  zero modes of other quasiprimaries. This has been checked for the Lee-Yang model \cite{Downing:2024nfb} and also for  the case of the ${\cal W}_3$ algebra \cite{Downing:2025huv}. We have extended these results now to a general chiral deformation, and the modular transformation is dictated by the iteration we have derived, involving the second order pole coefficient.

For the Ising model \cite{Downing:2023lnp}, and more recently for the symplectic fermions \cite{FaisalGerard}, the GGE and its modular transformations were given an interpretation in terms of a purely transmitting defect. The original GGE is interpreted as a defect wrapping the spatial direction of the cylinder, while after the modular S-transformation, the line defect is  along the time direction. In terms of figure \ref{fig: Figure from 2021}, the defect can be considered to lie along the red cycle; in the original picture, time runs along the green cycle and the defect is inserted in the trace over the cylinder Hilbert space. After the modular transformation, time now runs along the red cycle, the defect is therefore in the time-like direction, the Hilbert space has changed because of the presence of the defect and the  Hamiltonian is modified by the presence of the defect. The problem of modular transformation then is equivalent to a determination of the defect Hilbert space and defect Hamiltonian.}
Our results are independent of such an interpretation, and are more general in that it is applicable to any chiral deformation of the conformal theory. It would nevertheless be interesting to interpret the results of the modular transformation from this point of view. 

The particular identification as the original CFT with a single defect is also not the only interpretation of the modular transformed system - alternatively, it can be thought of as the oroginal CFT with the insertion of any number of suitably defined defects, or indeed simply as a bulk theory with a different dispersion relation. The essential point is that the modular transform can also be calculated by propagating a Hamiltonian along the time direction of the transformed torus.


One of the key assumptions of the proof was that result of the modular transformation could be expressed as an exponential the leading terms in $\tau$. These leading terms correspond to the nested products \eqref{Wnrecursionv2} which are in one-to-one correspondence with binary rooted trees, as explained in \cite{Downing:2025huv}. 
Furthermore, we have now found the multiplicity with which these should be counted -- so that the recursive form is not actually needed -- which we give in appendix \ref{sec:countingtrees}. 
If we consider these terms as connected trees (which they are), then since the counting includes a factor $1/|Aut(T)|$, where $Aut(T)$  is the automorphism group of the tree, then the exponentiation property just corresponds to summing over all trees, and not just connected trees. If this result was derived from an action with the leading terms being connected diagrams then we do not need to do any more to prove the exponential form -- but unfortunately we are not in the position. The relation to actions given in \cite{Dijkgraaf:1996iy} only works for sets of fields satisfying a pre-Lie algebra property which our more general expressions do not. There is, however, a good argument for why the exponentiation should hold.
While the calculation of the leading terms is a rigorous mathematical argument, the argument for the exponentiation is instead a physics argument.
In light of the interpretation of the modular transform as a defect Hamiltonian $H_D$ propagating along the defect, exponentiation is immediate\footnote{We thank I. Runkel for this observation.}.
If, for simplicity, we take $\tau = iL$, then 
it is clear that the result of rotating the plane should be of the form 
\begin{align}
    \langle e^{\alpha W_0} \rangle_{-1/\tau}
    = 
    \Tr_{\mathcal H}(e^{-L H_D})
    = \Tr_{\mathcal H}(e^{-L H_0 - L H'})
    \;,
\end{align}
which is only consistent if $H_0=2\pi L_0$, and $H'$ is independent of $L$.
In this way, we see that the insertion
$e^{-L H'}$ is entirely determined $H'$, that is by the order $L = - i\tau$ term, ie by the $O(\tau)$ term that we have calculated. We give a few more details of this argument in appendix \ref{app:expham}.
We hope to return to this again with a mathematical proof, but we still regard the theorem as proven, albeit partly using physics arguments.

A straightforward generalization of our results would be to include both chiral and anti-chiral deformations simultaneously,  which would seem natural from the point of view of GGEs.

Another direction to pursue would be to generalize our results and study the modular transformation of torus correlation functions in the presence of chiral deformations. For instance in \cite{Addabbo:2024ljn}, such mixed correlators have been studied for the so called heavy Heisenberg algebras. There are some general results that show  that the mixed correlators containing both zero modes and vertex operators transform as quasi-Jacobi forms. It would be interesting to extend our methods to this case. 

It is also important to remember that our results are obtained in an asymptotic expansion in the fugacities. A broader  understanding of how to derive the modular transformation of the deformed conformal field theory, without working term by term in the asymptotic expansion, remains an outstanding problem. In the two cases of the Ising model and the symplectic fermion,  the modular transformation could be calculated exactly. These are, of course, essentially free field theories and it would again be interesting to see if there are other models in which this can be done.

\section*{Acknowledgements}

We would like to thank 
Roger Behrend, Diptarka Das, Max Downing, Suresh Govindarajan, Sachin Grover, Faisal Karimi, Alok Laddha, Arkajyoti Manna, Sridip Pal, Ingo Runkel, 
Senan Sekhon, Shashank Sengar, Ronak M. Soni, and Jan Troost for many helpful discussions. We are especially thankful to Max Downing, Ronak M. Soni, and Jan Troost for a careful reading of the manuscript and for valuable feedback. 
TS is partially supported by a grant to CMI from the Infosys Foundation.

\begin{appendix}

\section{List of Integrals}
\label{listofintegrals}

We define the generalized weight-$k$ Weierstrass functions as follows: 
\be 
 {\cal P}_k(u) = \frac{(2\pi i)^k}{(k-1)!}\sum_{n\ne 0} \frac{n^{k-1} z^k}{1-q^n}  ~,
\ee 
where $q=e^{2\pi i \tau}$ and $z=e^{2\pi i u}$. This summation is valid when $|q| < |z| < 1$. We summarize the results of a few integrals that recur in our analysis:
    \begin{align}
        \int_{1}^{q}\frac{dz_i}{z_i}\mathcal{P}_{1}\Big(\frac{z_k}{z_i}\Big) &=(2\pi i)(i\pi + 2\pi i(u_k-\tau)) \label{intP1} \\
        \int_{1}^{q}\frac{dz_i}{z_i}\mathcal{P}_{2}\Big(\frac{z_k}{z_i}\Big) &=(2\pi i)^2 \label{intP2} \\
        \int_{1}^{q}\frac{dz_i}{z_i}\mathcal{P}_{m}\Big(\frac{z_k}{z_i}\Big)&=0~,~~m>2 \label{intPm>2}\\
     \int_{1}^{q}\frac{dz_k}{z_k}g^1_{m+1}\Big(\frac{z_i}{z_k}\Big) &=-\frac{(2\pi i)^2}{m}\mathcal{P}_{m}(z_i)~,~~~m>0
     \label{intg1}\\  \int_1^q \frac{dz_k}{z_k}\, u_k \, {\cal P}_{m+1}\left(\tfrac{z_i}{z_k}\right) &= (2\pi i)\Big [(i\pi + 2\pi i u_i)\delta_{m,1} + i\pi \delta_{m,2} - \frac{\tau}{m} {\cal P}_m(z_i)\Big] ~,~~m>0 \label{intuP}\\
        \int d\mu_{23} \, g^{1}_{m+1}\Big(\frac{z_4}{z_3}\Big)
        &= \int_{1}^{q}d\mu_{23}\, \mathcal{P}_{1}\Big(\frac{z_3}{z_2}\Big)\mathcal{P}_{m+1}\Big(\frac{z_4}{z_3}\Big)-\frac{ (2\pi i)^{4}}{2}\delta_{m,2}\cr
        &\hspace{1.2cm}-\left[\frac12(2\pi i)^4 +(2\pi i)^2\int_{1}^{q}\frac{dz_3}{z_3}\mathcal{P}_{1}\Big(\frac{z_4}{z_3}\Big)\right]\delta_{m,1}~,~~m>0
        \label{intg1P1m}
\end{align}
We show how the last equation is derived in detail. Let us evaluate the LHS of \eqref{intg1P1m}, assuming $m>0$:
 \begin{equation}\label{checkg1}
 \begin{aligned}
         \int d\mu_{23} \, g^{1}_{m+1}\Big(\frac{z_4}{z_3}\Big) =& \ (2\pi i)\tau \int \frac{dz_3}{z_3}  g^{1}_{m+1}\Big(\frac{z_4}{z_3}\Big) \\ =& \ -(2\pi i)^3\frac{\tau}{m} {\cal P}_m(z_4)
 \end{aligned}
 \end{equation}
where we used \eqref{intg1}. Now for the RHS of \eqref{intg1P1m}:
  \begin{equation}
 \begin{aligned}
        \int_{1}^{q}d\mu_{23}\, \mathcal{P}_{1}\Big(\frac{z_3}{z_2}\Big)\mathcal{P}_{m+1}\Big(\frac{z_4}{z_3}\Big)-\frac{ (2\pi i)^{4}}{2}\delta_{m,2}-\left[\frac12(2\pi i)^4 +(2\pi i)^2\int_{1}^{q}\frac{dz_3}{z_3}\mathcal{P}_{1}\Big(\frac{z_4}{z_3}\Big)\right]\delta_{m,1} \\ = \int_{1}^{q}\frac{dz_3}{z_3}(2\pi i) (i\pi - 2\pi i\tau +2\pi iu_3)\mathcal{P}_{m+1}\Big(\frac{z_4}{z_3}\Big)-\frac{ (2\pi i)^{4}}{2}\delta_{m,2}- (2\pi i)^4 (u_4+1-\tau)\delta_{m,1}
 \end{aligned}
 \end{equation}
 where we applied \eqref{intP1} to the first and last terms. Now, we assume $m>0$ and apply \eqref{intP2},\eqref{intPm>2} and \eqref{intuP}
to arrive at:
\begin{equation}
 \begin{aligned}
        \int_{1}^{q}d\mu_{23}\mathcal{P}_{1}\Big(\frac{z_3}{z_2}\Big)\mathcal{P}_{m+1}\Big(\frac{z_4}{z_3}\Big) &-\frac{ (2\pi i)^{4}}{2}\delta_{m,2}-\left[\frac12(2\pi i)^4 +(2\pi i)^2\int_{1}^{q}\frac{dz_3}{z_3}\mathcal{P}_{1}\Big(\frac{z_4}{z_3}\Big)\right]\delta_{m,1} \\ &= \frac{(2\pi i)^4}{2} (1-2\tau)  \delta_{m,1}  -\frac{ (2\pi i)^{4}}{2}\delta_{m,2}- (2\pi i)^4 (u_4+1-\tau)\delta_{m,1} \\ & \hspace{0.5cm} + \ (2\pi i)^3 \Big [(i\pi + 2\pi i u_4)\delta_{m,1} + i\pi \delta_{m,2} - \frac{\tau}{m} {\cal P}_m(z_4)\Big]  \\  &= -(2\pi i)^3 \frac{\tau}{m} {\cal P}_m(z_4) 
 \end{aligned}
 \end{equation}
 which agrees with what we got in \eqref{checkg1}.

\section{Useful Identities}
\label{realLemma}

We recall the definition of square modes of vertex operators (for more details we refer to \cite{zhu:1990, zhu:1996, Tuite}). This involves defining the vertex operator:
\be
V[A,u]:=V(e^{2\pi i u L_0} \tilde A, e^{2\pi i u} -1)~.
\ee 
The formal expansion of $V[A,u]$ in the periodic coordinate $u$ defines the square modes:
\be 
V[A,u] = \sum_{n\in \mathbb{Z}}\frac{ \tilde A[n]}{u^{n+1}}
\ee 
In the main text we have related the action of square modes on operators to OPE coefficients (see equation \eqref{nthorderpole}):  
\be 
\widetilde{(AB)}_n = (2\pi i)^{n}\, (\tilde A[n-1]\tilde{B})~.
\label{nthorderpoleAPP}
\ee 
We now use this to find simple relations between operators defined by the action of square modes. 

\subsection{Operator Identities}

Commutativity of the operators in the meromorphic conformal field theory means that the OPE coefficients satisfy simple identities. For instance we have 
   \begin{equation}
 (AB)_n(u)=   \sum_{j\ge0}(-1)^{n+j}\frac{1}{j!}\partial^j(BA)_{n+j}(u)~.
\end{equation}
For the vertex operators that we use to compute the torus correlators, this means that 
   \begin{equation}
 (2\pi i)^{h_{(AB)_n}}V( e^{2\pi i vL_0}\widetilde{(AB)}_n , e^{2\pi i v})=   \sum_{j\ge0}(-1)^{n+j}\frac{(2\pi i)^{h_{(BA)_{n+j}}}}{j!}\partial^j_u  V( e^{2\pi i vL_0}\widetilde{(BA)}_{n+j} , e^{2\pi i v})
\end{equation}
Using the relation we proved in \eqref{nthorderpoleAPP} we have: 
\begin{equation}
   (2\pi i)^{h_{(BA)_n}} V( e^{2\pi i vL_0}\widetilde{(BA)}_n , e^{2\pi i v}) = (2\pi i)^{h_A+h_B } V(e^{2\pi i v L_0} \tilde B[n-1]\tilde A, e^{2\pi i v})  ~,
\end{equation}
and a similar relation for $\widetilde{(AB)_n}$. Combining these equations, we find the relation between composite operators built out of square mode actions, in which the order of the operators has been flipped: 
   \begin{equation}
   \begin{aligned}
     V( e^{2\pi i vL_0}\tilde A[n-1]\tilde B , e^{2\pi i v})=   \sum_{j\ge0}\frac{(-1)^{n+j}}{j!}\partial^j_u  V( e^{2\pi i vL_0}\tilde B[n+j-1]\tilde A , e^{2\pi i v})
   \end{aligned}
\end{equation}
Recalling the definition of the torus correlators in \eqref{toruscorrelator}, one thereby obtains operator relations that are true within correlation functions on the torus ($z=e^{2\pi i u}$ as usual): 
\begin{align}
\langle\ldots ( A[m] B)(z)\ldots \rangle 
&=  \sum_{j\ge0}\frac{(-1)^{m+j+1}}{j!}\partial^j_u \langle \ldots ( B[m+j] A)(z) \ldots \rangle ~.
\label{eq:OpeDiff}
\end{align}
For $m=0$ we have
\begin{equation}
    \begin{aligned}
      \langle\ldots   ( A[0] B)(z) \ldots \rangle &=-
        \langle\ldots (B[0] A)(z) \ldots \rangle +    \langle\ldots \partial_{u} ( B[1] A)(z) \ldots \rangle + \text{Higher derivatives}\\
      \Rightarrow   \langle\ldots  ( W[0] W)(z) \ldots \rangle  &=       \frac{1}{2}\partial_{u}  \langle\ldots (W[1]W)(z)\ldots \rangle + \text{Higher derivatives}
    \end{aligned}
    \label{AOBflip}
\end{equation} 
In the second relation, we have shown that when $A=B=W$ , the operator obtained by the square mode action $ W[0] W$ is a total derivative. For $m=1$, on the other hand, we have 
\begin{equation}
    \begin{aligned}
    \langle \ldots  ( A[1] B)(z) \ldots\rangle  &=\langle \ldots  ( B[1] A)(z) \ldots \rangle - \partial_{u}  \langle \ldots ( B[2] A)(z) \ldots \rangle + \text{Higher  derivatives}\\
    \end{aligned}
    \label{A1Bflip}
\end{equation} 

\subsubsection{The Jacobi Identity}

Through standard contour arguments, one can obtain the Jacobi identity \cite{zhu:1990}:
\be
\label{eq:zhusquare}
(b[n]a)[m]=\sum_{i}\binom{n}{i}\Big((-1)^ib[n-i]a[m+i]-(-1)^{n+i}a[n+m-i]b[i]\Big)~.
\ee
A consequence of the Jacobi identity that will prove useful is
\begin{align}
     (X[0]W)[n+1]+(X[1]W)[n]&=X[1]W[n]-W[n]X[1] 
\end{align}
which can be arrived at by simply summing up after expanding the terms on the LHS using the Jacobi identity. Two identities that follow from this relation will be repeatedly used in the text, so we collect them below: 
\begin{align}
\label{W0W1}
     (W[0]W)[1] &=-\frac12 (W[1]W)[0]\\
     (W[0]W)[2]&=-(W[1]W)[1] ~.
\label{W0W2}
\end{align}

\subsection{Correlator Identities}

\subsubsection{Periodicity properties of correlators with zero-modes} \label{periodicity}

As we have discussed  in the introduction, the torus correlators of a meromorphic conformal field theory,  defined in \eqref{toruscorrelator}
are singled valued if we perform the shift in the $u$ variable by $1$ or $\tau$. However, in the presence of a zero mode insertion, this is no longer the case. Below we derive the periodicity property of torus correlators in the presence of a single zero mode: 
\begin{equation}
    \begin{aligned}
        \langle X_0\, X_2(qz_2)\ldots X_n(z_n)\rangle=&\Tr( X_0\, V((qz_2)^{L_0}\tilde X_2, qz_2)\ldots V(z_n^{L_0}\tilde X_n,z_n)q^{L_0-\frac{c}{24}})\\
        =&\Tr( X_0\, V(z_3^{L_0} \tilde X_3, z_3)\ldots V(z_n^{L_0}\tilde X_n, z_n)V((qz_2)^{L_0} \tilde X_2, qz_2)q^{L_0-\frac{c}{24}})\\
         =&\Tr( X_0\, V(z_3^{L_0}\tilde X_3, z_3)\ldots V(z_n^{L_0}\tilde X_n, z_n)q^{L_0-\frac{c}{24}}V(z_2^{L_0}\tilde X_2,z_2))\\
          =&\Tr( X_0 \, V(z_2^{L_0} \tilde X_2,z_2) V(z_3^{L_0}\tilde X_3,z_3)\ldots V(z_n^{L_0}\tilde X_n, z_n)q^{L_0-\frac{c}{24}})\\
          &-\Tr( [X_0,V(z_2^{L_0} \tilde X_2,z_2)]V(z_3^{L_0}\tilde X_3,z_3)\ldots V(z_n^{L_0}\tilde X_n,z_n)q^{L_0-\frac{c}{24}})\\
        \end{aligned}
\end{equation}   
The commutator can be replaced by the square mode action as follows \cite{Gaberdiel:2012yb}: 
\be 
[X_0, V(z^{L_0} \tilde Y,z)] =  2\pi i \, V(z^{L_0} \tilde X[0]\tilde Y,z)
\label{squaremodecomm}
\ee 
Thus reverting to the angular bracket notation for the torus correlator, we find the relation
\begin{multline}
\label{zeromodeperiodicity}
    \langle X_0\, X_2(qz_2)\ldots X_n(z_n)\rangle =\langle X_0\, X_2(z_2)\ldots X_n(z_n)\rangle\\
           -(2\pi i)\langle (X[0]X_2)(z_2)X_3(z_3)\ldots X_n(z_n)\rangle~.
\end{multline}

\subsection{Identities for Integrated Correlators}
\label{corridentities}

Let us recall the map of the operators from the cylinder to the plane that was defined in \eqref{operatormap}: 
\be 
 V(A ,u) \rightarrow  (2\pi i)^{h_A} V(e^{2\pi i u L_0} \tilde A, e^{2\pi i u}) ~.
\ee 
For primary operators, the $\tilde A$ operator coincides with the operator $A$ itself, while for quasiprimaries $\tilde A$ is different from $A$. The stress tensor is an example of a spin two quasiprimary, and if we denote the state by 
\be 
\omega = L_{-2}\Omega~,
\ee 
where $\Omega $ is the vacuum, it gets mapped onto a state which we denote $\tilde\omega$ (we follow the conventions of \cite{Tuite, Maloney:2018hdg})
\be 
\label{omegatildedefn}
\tilde\omega = L_{-2}\Omega - \frac{c}{24}\Omega~. 
\ee 
Now, the zero mode of $\omegat$ is given by (see \cite{Gaberdiel:2008pr, Maloney:2018hdg, Ashok:2024zmw} for more details)
\be 
\omegat[0]=\frac{1}{(2\pi i)}(L_0 + L_{-1}) = \frac{1}{2\pi i} \partial_u ~. 
\ee 
Let us consider an integrated correlator in which we consider the zero-mode insertion of the stress tensor:
\be 
{\cal C}=\int_{1}^{q}d\mu_{2n} \langle \omegat_0\,  X_1(z_1)\dots X_n(z_n)\rangle~.
\ee 
We now commute the zero mode through all the other fields in the trace, using \eqref{squaremodecomm}:
\be 
\langle \ldots [\omegat_0, X(z)] \ldots \rangle = 2\pi i\langle\ldots  (\omegat[0]X)(z)\ldots \rangle =  \langle \ldots \partial_u X(z) \ldots \rangle ~.
\ee 
Thus the integrated correlator can be written as: 
\begin{multline*}
{\cal C}=\int_{1}^{q}d\mu_{2n} \langle (\omegat[0]X_1)(z_1)\dots X_n(z_n)\rangle\\
+\sum_{m=2}^n\int_{1}^{q}d\mu_{2n}\langle X_1(z_1)\dots (\omegat[0]X_m)(z_m)\dots X_n(z_n)\rangle+{\cal C}~.
\end{multline*} 
Now the correlator cancels on both sides and we obtain the identity
\be
    \begin{aligned}        0=&\int_{1}^{q}d\mu_{2n}\Big[\langle (\omegat[0]X_1)(z_1)\dots X_n(z_n)\rangle+ \sum_{m=2}^n \partial_{u_m}\langle X_1(z_1) X_2(z_2)\dots X_m(z_m)\dots X_n(z_n)\rangle\Big]~.
\end{aligned}
\ee 
As the correlator is single-valued, the total derivatives in the second term leads to a vanishing result \footnote{Note that we did not need to exchange integrals to perform the $z_m$ integral, as the $u_j$ periodicity of the correlator is not spoiled by the $z_i$ integrals $(i<m)$.}. Thus, we obtain
\begin{equation}
   \int_{1}^{q}d\mu_{2n} \langle \partial_{u_1}X_1(z_1)\dots X_n(z_n)\rangle =0~.
\end{equation}
Since $z_1$ is unintegrated, one can rewrite this  as
\be 
\int_1^q d\mu_{1n} h(z_1) \langle \partial_{u_1}X_1(z_1)\dots X_n(z_n)\rangle =0~,
\ee 
for an arbitrary function $h(z)$. 
We make a few remarks about this result:
\begin{itemize}
    \item The fields $X_1, \dots, X_n$ are arbitrary and it is important to note that $X_1(z_1)$ is not integrated over. 
    \item There is nothing sacred about $X_1$ being in the first entry of the correlator; since we have a meromorphic conformal field theory, the total derivative  can be at any of the entries in the correlator; the single-valuedness of the correlator ensures that the resulting integrated correlator would still vanish. So we have a result which we shall refer to as our first Lemma. 
\end{itemize}
\underline{\bf Lemma 1}: 
    \begin{equation}
\int_{1}^{q}d\mu_{1n}\, h(z_j) \langle X_1(z_1)\dots \partial_{u_j}X_j(z_j)\ldots X_n(z_n)\rangle =0~.
        \label{lemma1}
\end{equation}

This result can be used to derive stronger constraints on more general integrated correlators. For instance, let us start with an integrated correlator with all but one $W$-insertions, and a zero mode insertion. We again move the zero mode through the trace: 
\begin{equation}
    \begin{aligned}
        {\cal C}=& \frac{1}{2\pi i}\int_{1}^{q}d\mu_{2n}\, \langle W_0\, X(z_1) W(z_2)\dots W(z_n)\rangle\\
        =&\int_{1}^{q}d\mu_{2n}\, \langle (W[0]X)(z_1)\dots W(z_n)\rangle\\
        &+\sum_{j=2}^{n}\int_{1}^{q}d\mu_{2n} \, \langle X(z_1)W(z_2)\ldots (W[0]W)(z_j)\ldots W(z_n)\rangle+{\cal C} ~. 
            \end{aligned}
\end{equation}
Cancelling the correlator that we began with, and using 
\be 
\langle \ldots W[0]W(z) \ldots \rangle = \frac12 \langle \ldots  \partial_u (W[1]W)(z) \ldots\rangle+ \text{Higher derivatives}~,
\ee 
we obtain the identity:
\begin{equation}
    \begin{aligned}
        0=&\int_{1}^{q}d\mu_{2n}\langle (W[0]X)(z_1)\dots W(z_n)\rangle\\
        &+\frac{ 1}{2}\sum_{j=2}^n\int_{1}^{q}d\mu_{2n}\, \partial_{u_j}\langle X(z_1)W(z_2)\ldots (W[1]W)(z_j)\dots W(z_n)\rangle~,\\
      \text{or}\quad   0=&\int_{1}^{q}d\mu_{2n}\,\langle (W[0]X)(z_1) W(z_2)\dots W(z_n)\rangle~.
    \end{aligned}
    \label{lemmaW0X}
\end{equation}
In the last equality we have again used the single-valuedness of the correlator to conclude that the total derivative term leads to a vanishing contribution for each $j$. One can add a further integral over $z_1$ to obtain the identity:
\be 
\int_{1}^{q}d\mu_{1n}\, h(z_1) \langle (W[0]X)(z_1) W(z_2)\dots W(z_n)\rangle = 0~.
\ee
Once again, the position of the operator $X$ is irrelevant and we obtain 

\underline{\bf Lemma 2:}
\be 
\label{lemma2}
\int_{1}^{q}d\mu_{1n}~ h(z_j)~\langle W(z_1)\ldots (W[0]X)(z_j)\dots W(z_n)\rangle = 0~.
\ee 
A related identity can be derived by 
using the general result (see \eqref{AOBflip}) 
\be 
\langle \ldots \big(W[0]X(z)+X[0]W(z)\big) \ldots \rangle = \langle \ldots  \partial_u (X[1]W)(z) \ldots \rangle+ \text{Higher derivatives}~.  
\ee 
Substituting this into the second lemma, we see that the derivative term leads to a vanishing contribution on account of Lemma 1. Thus we have 

\underline{\bf Lemma 3:}~
\be 
\int_{1}^{q}d\mu_{1n}~ h(z_j)~\langle W(z_1)\ldots (X[0]W)(z_j)\dots W(z_n)\rangle = 0~.
\label{lemmaXOW}
\ee

\section{Generalization to Local Fields}
\label{generalizationofZhuresult}

In this Appendix we show that our recursive result for the integrated $n$-point correlator in equation 
\eqref{npointaslowerpoint} actually holds for an arbitrary local operator $W(z)$, and is not restricted to operators with a fixed scaling dimension. In order to prove this point, we go back and examine the properties of $W(z)$ that were used in our derivation of the recursion.

\subsection{Locality}

A finite sum of local fields is, of course, a local field. By virtue of \eqref{nthorderpole}, the square modes $(\tilde A[n]\tilde B)$ are associated to the ${n+1}$st order pole from the OPE of $A(z)$ and $B(w)$. If $A(z)$ and $B(w)$ were linear combinations of other fields, such as $A(z)=\sum_i A^i(z)$ and $B(w) = \sum_j B^j(w)$, we would have:
\begin{equation}
   \langle \ldots A[n]B(w) \ldots \rangle =  \langle \ldots \sum_{i,j} A^i[n]B^j(w)\ldots \rangle 
    \label{AnBsplit}
\end{equation}
This is an identity we will apply repeatedly in the coming subsections. Another identity that we would like to use is:
\begin{equation}
   \langle \ldots [A_0,B(w)] \ldots \rangle = 2\pi i \langle \ldots (A[0]B)(w) \ldots \rangle
\end{equation}
which again follows just be expanding out either side using $A(z)=\sum_i A^i(z)$ and $B(w) = \sum_j B^j(w)$. 

Local fields in a meromorphic conformal field theory obey:
\begin{equation}
    A(z)B(w) = B(w)A(z)
\end{equation}
which is all that goes into the proof of relations we have used repeatedly in the derivation such as:
\begin{equation}
    \begin{aligned}
        \langle \ldots (A[0]B)(v) \ldots \rangle &=- \langle \ldots\Big[ (B[0]A)(v)- \partial_{v} (B[1]A)(v) \Big] \ldots \rangle + \text{Higher order derivatives}
    \end{aligned}
\end{equation}
Likewise, the Jacobi identity is simply a consequence of contour deformation arguments and the fact that the singularities of correlators of local operators are prescribed by the OPEs. 

\subsection{The Zhu formula}

A crucial tool in our analysis is the Zhu formula. It is clear that even if $W(z)$ were a sum of fields that are not of a definite $L[0]$ weight, one could decompose the correlator into correlators with definite weights, apply the Zhu formula and resum the terms. This is because the terms that appear in the Zhu formula are either zero modes of operators or OPE coefficients of pairwise chosen operators. Both of these can again easily be resummed back to expressions involving the local fields, for example, as shown above in \eqref{AnBsplit}. Let us see this a little more explicitly. If $X^i(z) := \sum_{j=1}^{J_i} G^{ij}(z)$, each of the $G$s being of definite weight, we have:
\begin{align}
        \langle X^1(z_1)\ldots X^n(z_n)\rangle &= \sum_{j_1,j_2,\dots j_n} \langle G^{1j_1}(z_1) \ldots G^{nj_n}(z_n)\rangle \cr
        &= \sum_{j_1,j_2,\dots j_n} \langle  G^{1j_1}_0\, \ldots G^{nj_n}(z_n)\rangle \\ &+ \sum_{j_1,j_2,\dots j_n} \sum_{k=2}^n \sum_{m=0}^\infty {\cal P}_{m+1}\Big(\frac{z_k}{z_1}\Big) \, \langle G^{2j_2}(z_2) \dots (G^{1j_1}[m]G^{kj_k})(z_k)\ldots G^{nj_n}(z_n)\rangle ~.\nonumber
\end{align}
which is the same as:
\begin{equation}
         \langle X^1_0\, X^2(z_2)\ldots X^n(z_n)\rangle  + \sum_{k=2}^n \sum_{m=0}^\infty {\cal P}_{m+1}\Big(\frac{z_k}{z_1}\Big) \, \langle X^0_0\,  X^2(z_2) \dots (X^1[m]X^k)(z_k)\ldots X^n(z_n)\rangle ~. 
\end{equation}
This can be shown simply re-expanding the latter in terms of the $G^{ij}$s. The same can be shown to be true for the generalized Zhu recursion involving zero mode insertions. 

\subsection{Periodicity}
Correlators (perhaps with a zero mode insertion) of local operators of definite weight have periodicity properties derived in \ref{periodicity}, and we made good use of this property in our derivation. Consider such a torus correlator:
\begin{equation}
        \langle X^0_0\, X^1(z_1)\ldots X^n(z_n)\rangle = \sum_{j_0,j_2,\dots j_n} \langle G^{0j_0}_0\,  G^{1j_1}(z_1) \ldots G^{nj_n}(z_n)\rangle 
\end{equation}
We can apply the periodicity result on each term on the RHS:
\begin{align}
        \langle X^0_0\, X^1(qz_1)\ldots X^n(z_n)\rangle &= \sum_{j_0,j_1,\dots j_n} \langle G^{0j_0}_0\,  G^{1j_1}(qz_1) \ldots G^{nj_n}(z_n)\rangle \cr
        &= \sum_{j_0,j_1,\dots j_n} \langle G^{0j_0}_0\, G^{1j_1}(z_1) \ldots G^{nj_n}(z_n)\rangle \\
        &\hspace{3cm}- 2\pi i \sum_{j_0,j_1,\dots j_n} \langle (G^{0j_0}[0]G^{1j_1})(z_1) \ldots G^{nj_n}(z_n)\rangle \cr &=  \langle X^0_0\, X^1(z_1)\ldots X^n(z_n)\rangle -2\pi i \langle (X^0[0]X^1)(z_1) X^2(z_2) \cdots X^n(z_n) \rangle\nonumber ~.
\end{align}
If the zero mode was just that of the identity operator, the second term drops out and we recover the periodicity property of a correlator of local operators. 

We see that every property we assume for $W(z)$ is also one that arbitrary local fields satisfy. We infer that the recursive formula we have derived for the integrated correlator in \eqref{npointaslowerpoint} is in fact valid for an arbitrary local field $W(z)$.

\section{Counting of trees}
\label{sec:countingtrees}

As was shown in \cite{Downing:2025huv}, the individual terms in the nested products \eqref{Wnrecursionv2} are in one-to-one correspondence with binary rooted trees, for example: 
\newcommand{\lb}{)^{\vphantom{y}}_2}
\tikzset{
        blank/.style={draw=none},
         edge from parent/.style=
         {draw,edge from parent path={(\tikzparentnode) -- (\tikzchildnode)}},
         level distance=1.cm}
\[
(WW\lb  = \!\!\!\!
\raisebox{-12mm}{ 
    \begin{tikzpicture}
    \node at (-.4,0.0) {$W$};
    \node at (.4,0.0) {$W$};
    \fill (0,1) circle (2.4pt);
    \draw (0,1) -- (-.4,0.2);
    \draw (0,1) -- (.4,0.2);
\end{tikzpicture}}
,\;\,
(W(WW\lb\lb = \!\!\!\! 
   \raisebox{-2.cm}{
    \begin{tikzpicture}
        \node at (-.7,0.0) {$W$};
    \node at (.3,-.8) {$W$};
    \node at (1.1,-.8) {$W$};
    \fill (0,1) circle (2.4pt);
    \draw (0,1) -- (-.7,0.2);
    \draw (0,1) -- (.7,0.2);
\fill (.7,.2) circle (2.4pt);
    \draw (.7,.2) -- (1.1,-0.6);
    \draw (.7,.2) -- (.3,-0.6);
\end{tikzpicture}}
\!\!\!\!,\;\,
((WW\lb (WW)\lb\lb = \!\!\!\!\!\!\!
   \raisebox{-2.cm}{
    \begin{tikzpicture}
    \node at (.3,-.8) {$W$};
    \node at (1.1,-.8) {$W$};
    \node at (-.3,-.8) {$W$};
    \node at (-1.1,-.8) {$W$};
    \fill (0,1) circle (2.4pt);
    \draw (0,1) -- (-.7,0.2);
    \draw (0,1) -- (.7,0.2);
\fill (-.7,.2) circle (2.4pt);
    \draw (-.7,.2) -- (-1.1,-0.6);
    \draw (-.7,.2) -- (-.3,-0.6);
\fill (.7,.2) circle (2.4pt);
    \draw (.7,.2) -- (1.1,-0.6);
    \draw (.7,.2) -- (.3,-0.6);
\end{tikzpicture}}
\;.
\]
Because the product $(XY)_2$ is commutative (although not associative), each tree has many different planar realisations:
\[
(W(WW\lb\lb = \!\!\!\! 
   \raisebox{-2.cm}{
    \begin{tikzpicture}
        \node at (-.7,0.0) {$W$};
    \node at (.3,-.8) {$W$};
    \node at (1.1,-.8) {$W$};
    \fill (0,1) circle (2.4pt);
    \draw (0,1) -- (-.7,0.2);
    \draw (0,1) -- (.7,0.2);
\fill (.7,.2) circle (2.4pt);
    \draw (.7,.2) -- (1.1,-0.6);
    \draw (.7,.2) -- (.3,-0.6);
\end{tikzpicture}}
=
((WW\lb W\lb = \!\!\!\!\!\!\!
   \raisebox{-2.cm}{
    \begin{tikzpicture}
    \node at (.7,0) {$W$};
    \node at (-.3,-.8) {$W$};
    \node at (-1.1,-.8) {$W$};
    \fill (0,1) circle (2.4pt);
    \draw (0,1) -- (-.7,0.2);
    \draw (0,1) -- (.7,0.2);
\fill (-.7,.2) circle (2.4pt);
    \draw (-.7,.2) -- (-1.1,-0.6);
    \draw (-.7,.2) -- (-.3,-0.6);
\end{tikzpicture}}
\;.
\]
We have now found a formula for the total number of equivalent terms that appear in the sum \eqref{Wnrecursionv2}. If we consider any particular realisation of a tree with $N$ internal nodes indexed by $i$, then the total number $N(T)$ of equivalent terms is
\begin{equation}
\label{eq:N(T)}
N(T) = \frac{
2^N}{|Aut(T)|}\,\frac{N!}{\prod_i n_i} 
\;,
\end{equation}
where $n_i$ is the number of internal nodes in the tree rooted at the $i$th node and the automorphism group has a factor of $\mathbb Z_2$ for each internal node with identical trees hanging from it. For example, if we consider the tree (with labelled nodes)
\[
((WW\lb (W(WW)\lb\lb\lb = \!\!\!\!\!\!\!
   \raisebox{-2.9cm}{
    \begin{tikzpicture}
    \node at (.3,-.8) {$W$};
    \node at (-.3,-.8) {$W$};
    \node at (-1.1,-.8) {$W$};
    \node at (1.35,-1.6) {$W$};
    \node at (0.85,-1.6) {$W$};
    \fill (0,1) circle (2.4pt);
    \draw (0,1) -- (-.7,0.2);
    \draw (0,1) -- (.7,0.2);
\fill (-.7,.2) circle (2.4pt);
    \draw (-.7,.2) -- (-1.1,-0.6);
    \draw (-.7,.2) -- (-.3,-0.6);
\fill (.7,.2) circle (2.4pt);
    \draw (.7,.2) -- (1.1,-0.6);
    \draw (.7,.2) -- (.3,-0.6);
\fill (1.1,-.6) circle (2.4pt);
    \draw (1.1,-0.6) -- (1.35,-1.4);
    \draw (1.1,-0.6) -- (0.85,-1.4);
    \node at (0.3,1) {$1$};
    \node at (-1,.2) {$2$};
    \node at (1.1,.2) {$3$};
    \node at (1.4,-.6) {$4$};
\end{tikzpicture}}
\]
then there are 4 internal nodes, $N=4$, $n_1=4$, $n_2=1$, $n_3=2$ and $n_4=1$. The automorphism group has a $\mathbb Z_2$ for the nodes 2 and 4 and so $|Aut(T)|=4$ and
\[
N(T) = \frac{2^4}{4}\frac{ 4!}{(4 \cdot 2 \cdot 1 \cdot 1)}
= 12\;.
\]
This agrees with the factor in equation (C.10) in \cite{Downing:2025huv} and it is easy to check that this formula reproduces all the factors in (C.6)--(C.11).

The explanation for the formula \eqref{eq:N(T)} is very simple. Each permutation of $N$ elements corresponds to a rooted binary tree, and each distinct planar realisation appears $N!/(\prod_i n_i)$ times (see e.g.
\cite{JonesYeats} which also makes a connection to quantum field theory). 
There are $2^N/|Aut(T)|$ distinct planar realisations of any given tree, hence we arrive at $N(T)$.

\section{The defect Hamiltonian}
\label{app:expham}

The argument that the modular transform exponentiates and is determined entirely by the order $O(\tau)$ term comes from the interpretation of the torus functions as the results of propagating along the time direction by a Hamiltonian that does itself not depend on the length of the time direction, so that, in general, if the torus is of sides of lengths $R$ and $L$ with $\tau = i L/R$, then 
\begin{equation}
Z = \Tr_{\cH_1}( e^{-R H_1(L)} ) = \Tr_{\cH_2}( e^{-L H_2(R)} )
\label{eq:gen}
\end{equation}
where $H_1$ is the Hamiltonian acting on the Hilbert space $\cH_1$, and similarly $H_2$ is the Hamiltonian in the modular transformed channel acting on the space $\cH_2$. 

We now compare the left hand side of \eqref{eq:gen} with the left hand side of \eqref{OGeqn}. We see
\begin{align}
    - R H_1(L) = - 2\pi\frac{R}{L}(L_0 - c/24) + \frac{\alpha}{\tau^w}\, W_0\;.
\end{align}
For this to be true, we need 
\begin{align}
    H_1(L) 
    = \frac{2\pi}{L}(L_0 - c/24) - \frac{\alpha}{R\tau^w}\, W_0
    = \frac{2\pi}{L}(L_0 - c/24) - \frac{\alpha}{R(iL/R)^w}\, W_0\;.
\end{align}
and hence 
\begin{align}
    \alpha = \tilde\alpha (-i)^w R^{1-w}~,
\end{align}
for some dimensionless constant $\tilde\alpha$, and 
\begin{align}
    H_1(L) 
    = \frac{2\pi}{L}(L_0 - c/24) - \frac{\tilde\alpha}{L^w}\, W_0.
\end{align}
Now looking at the right-hand side of \eqref{eq:gen}, we see that if $H_2(R)$ is only a function of $\tilde\alpha$ and $R$, then we must have the $L$ dependence arising from 
\begin{align}
    - L H_2(R) 
    =\frac{iL}R i R H_2(R)
    = \tau (i R H_2(R))~,
\end{align}
and that this is determined purely by the order $\tau$ terms in the modular transform, and the full modular transform is given exactly by the exponential of the $O(\tau)$ term.

\end{appendix}


\begin{thebibliography}{99}

\bibitem{Belavin:1984vu}
A.~A.~Belavin, A.~M.~Polyakov and A.~B.~Zamolodchikov,
``Infinite Conformal Symmetry in Two-Dimensional Quantum Field Theory,''
Nucl. Phys. B \textbf{241} (1984), 333-380
doi:10.1016/0550-3213(84)90052-X

\bibitem{Zamolodchikov:1990ji}
A.~B.~Zamolodchikov and A.~B.~Zamolodchikov,
``Conformal field theory and 2-D critical phenomena. 6. Modular bootstrap,''
ITEP-90-103.


\bibitem{Cardy:1986ie}
J.~L.~Cardy,
``Operator Content of Two-Dimensional Conformally Invariant Theories,''
Nucl. Phys. B \textbf{270} (1986), 186-204
doi:10.1016/0550-3213(86)90552-3

\bibitem{Dijkgraaf:1996iy}
R.~Dijkgraaf,
``Chiral deformations of conformal field theories,''
Nucl. Phys. B \textbf{493} (1997), 588-612
doi:10.1016/S0550-3213(97)00153-3
[arXiv:hep-th/9609022 [hep-th]].

\bibitem{Sasaki:1987mm}
R.~Sasaki and I.~Yamanaka,
``Virasoro Algebra, Vertex Operators, Quantum {Sine-Gordon} and Solvable Quantum Field Theories,''
Adv. Stud. Pure Math. \textbf{16} (1988), 271-296
RRK-87-3.

\bibitem{Eguchi:1989hs}
T.~Eguchi and S.~K.~Yang,
``Deformations of Conformal Field Theories and Soliton Equations,''
Phys. Lett. B \textbf{224} (1989), 373-378
doi:10.1016/0370-2693(89)91463-9




\bibitem{Bazhanov:1994ft}
V.~V.~Bazhanov, S.~L.~Lukyanov and A.~B.~Zamolodchikov,
``Integrable structure of conformal field theory, quantum KdV theory and thermodynamic Bethe ansatz,''
Commun. Math. Phys. \textbf{177} (1996), 381-398
doi:10.1007/BF02101898
[arXiv:hep-th/9412229 [hep-th]].

\bibitem{Fioravanti:1995cq}
D.~Fioravanti, F.~Ravanini and M.~Stanishkov,
``Generalized KdV and quantum inverse scattering description of conformal minimal models,''
Phys. Lett. B \textbf{367} (1996), 113-120
doi:10.1016/0370-2693(95)01463-2
[arXiv:hep-th/9510047 [hep-th]].


\bibitem{Downing:2021mfw}
M.~Downing and G.~M.~T.~Watts,
``Free fermions, KdV charges, generalised Gibbs ensembles and modular transforms,''
JHEP \textbf{06} (2022), 036
doi:10.1007/JHEP06(2022)036
[arXiv:2111.13950 [hep-th]].

\bibitem{Downing:2023lop}
M.~Downing,
``Modular transform of free fermion generalised Gibbs ensembles and generalised power partitions,''
[arXiv:2310.07601 [hep-th]].

\bibitem{Downing:2023lnp}
M.~Downing and G.~M.~T.~Watts,
``Free fermions, KdV charges, generalised Gibbs ensembles, modular transforms and line defects,''
JHEP \textbf{01} (2024), 041
doi:10.1007/JHEP01(2024)041
[arXiv:2311.04564 [hep-th]].


\bibitem{Downing:2024nfb}
M.~Downing and F.~Karimi,
``Modular Properties of Generalised Gibbs Ensembles,''
SciPost Phys. \textbf{18} (2025), 085
doi:10.21468/SciPostPhys.18.3.085
[arXiv:2410.06288 [hep-th]].


\bibitem{Zamolodchikov:1985wn}
A.~B.~Zamolodchikov,
``Infinite Additional Symmetries in Two-Dimensional Conformal Quantum Field Theory,''
Theor. Math. Phys. \textbf{65} (1985), 1205-1213
doi:10.1007/BF01036128


\bibitem{Kupershmidt:1989bf}
B.~A.~Kupershmidt and P.~Mathieu,
``Quantum Korteweg-de Vries Like Equations and Perturbed Conformal Field Theories,''
Phys. Lett. B \textbf{227} (1989), 245-250
doi:10.1016/S0370-2693(89)80030-9


\bibitem{Bazhanov:2001xm}
V.~V.~Bazhanov, A.~N.~Hibberd and S.~M.~Khoroshkin,
``Integrable structure of W(3) conformal field theory, quantum Boussinesq theory and boundary affine Toda theory,''
Nucl. Phys. B \textbf{622}, 475-547 (2002)
doi:10.1016/S0550-3213(01)00595-8
[arXiv:hep-th/0105177 [hep-th]].

\bibitem{Ashok:2024zmw}
S.~K.~Ashok, S.~Parihar, T.~Sengupta, A.~Sudhakar and R.~Tateo,
``Integrable structure of higher spin CFT and the ODE/IM correspondence,''
JHEP \textbf{07} (2024), 179
doi:10.1007/JHEP07(2024)179
[arXiv:2405.12636 [hep-th]].

\bibitem{Ashok:2024ygp}
S.~K.~Ashok, S.~Parihar, T.~Sengupta, A.~Sudhakar and R.~Tateo,
``Thermal correlators and currents of the $ {\mathcal{W}}_3 $ algebra,''
JHEP \textbf{01} (2025), 154
doi:10.1007/JHEP01(2025)154
[arXiv:2410.11748 [hep-th]].

\bibitem{Downing:2025huv}
M.~Downing, F.~Karimi, T.~Sengupta, A.~Sudhakar and G.~M.~T.~Watts,
``Modular Properties of $\mathcal{W}_3$ Generalised Gibbs Ensembles,''
[arXiv:2508.16258 [hep-th]].


\bibitem{Gaberdiel:2012yb}
M.~R.~Gaberdiel, T.~Hartman and K.~Jin,
``Higher Spin Black Holes from CFT,''
JHEP \textbf{04} (2012), 103
doi:10.1007/JHEP04(2012)103
[arXiv:1203.0015 [hep-th]].

\bibitem{Gaberdiel:2008pr}
M.~R.~Gaberdiel and C.~A.~Keller,
``Modular differential equations and null vectors,''
JHEP \textbf{09} (2008), 079
doi:10.1088/1126-6708/2008/09/079
[arXiv:0804.0489 [hep-th]].


\bibitem{Kraus:2011ds}
P.~Kraus and E.~Perlmutter,
``Partition functions of higher spin black holes and their CFT duals,''
JHEP \textbf{11} (2011), 061
doi:10.1007/JHEP11(2011)061
[arXiv:1108.2567 [hep-th]].


\bibitem{Iles:2013jha}
N.~J.~Iles and G.~M.~T.~Watts,
``Characters of the $W_3$ algebra,''
JHEP \textbf{02} (2014), 009
doi:10.1007/JHEP02(2014)009
[arXiv:1307.3771 [hep-th]].

\bibitem{Iles:2014gra}
N.~J.~Iles and G.~M.~T.~Watts,
``Modular properties of characters of the W$_{3}$ algebra,''
JHEP \textbf{01} (2016), 089
doi:10.1007/JHEP01(2016)089
[arXiv:1411.4039 [hep-th]].
\bibitem{FaisalGerard}
F.~Karimi and G.~M.~T.~Watts,
``Modular Properties of Symplectic Fermion Generalised Gibbs Ensemble,''
[arXiv:2603.19383 [hep-th]].


\bibitem{Gaberdiel:1994fs}
M.~Gaberdiel,
``A General transformation formula for conformal fields,''
Phys. Lett. B \textbf{325} (1994), 366-370
doi:10.1016/0370-2693(94)90026-4
[arXiv:hep-th/9401166 [hep-th]].


\bibitem{zhu:1990}
Y. Zhu,
  ``Vertex operator algebras, elliptic functions and modular forms," Ph.D. dissertation,
Yale Univ., 1990. https://api.semanticscholar.org/CorpusID:117186653

\bibitem{zhu:1996}
Y. Zhu, 
``Modular Invariance of Characters of Vertex Operator Algebras," Journal of the American Mathematical Society, Vol. 9, No. 1 (Jan., 1996), pp. 237-302







\bibitem{Goddard:1989dp}
P.~Goddard, 
``Meromorphic Conformal Field Theory,'' in: Infinite Dimensional Lie Algebras and Lie Groups, edited by V.~Kac, page 556, World Scientific, Singapore, New Jersey, Hong Kong, DAMTP-89-01.

\bibitem{Gaberdiel:1999mc}
M.~R.~Gaberdiel,
``An Introduction to conformal field theory,''
Rept. Prog. Phys. \textbf{63} (2000), 607-667
doi:10.1088/0034-4885/63/4/203
[arXiv:hep-th/9910156 [hep-th]].

\bibitem{AM}
A.~A.~Albert, B.~Muckenhoupt, ``On matrices of trace zero'', 
Michigan Math. J. 4 (1), 1-3, (1957),
DOI: 10.1307/mmj/1028990168.


\bibitem{Addabbo:2024ljn}
D.~Addabbo and C.~A.~Keller,
``Modularity of vertex operator algebra correlators with zero modes,''
J. Algebra \textbf{692} (2026), 27-69
doi:10.1016/j.jalgebra.2025.11.028
[arXiv:2411.08008 [math.QA]].




\bibitem{Tuite}
M.~P.~Tuite,
  ``Modular Forms in Vertex Operator Algebras,'' \url{https://legacy.slmath.org/attachments/sgw/449/tuite.pdf}.

  
\bibitem{Maloney:2018hdg}
A.~Maloney, G.~S.~Ng, S.~F.~Ross and I.~Tsiares,
``Thermal Correlation Functions of KdV Charges in 2D CFT,''
JHEP \textbf{02} (2019), 044
doi:10.1007/JHEP02(2019)044
[arXiv:1810.11053 [hep-th]].



\bibitem{JonesYeats}
B.\ R.~Jones and K.~Yeats,
``Tree hook length formulae, Feynman rules and B-series,''
Ann. Inst. Henri Poincaré Comb. Phys. Interact. 2 (2015), 413,
doi:10.4171/AIHPD/22
[arXiv:1412.6053]


\end{thebibliography}
\end{document}